\documentclass[useAMS,usenatbib]{mn2e}
\pdfoutput=1

\newcommand\scale{0.2}

\usepackage{changepage}
\usepackage{booktabs}
\usepackage{aas_macros}
\usepackage{epsfig}
\usepackage{amsmath}
\usepackage{amssymb}
\usepackage{comment}
\usepackage{caption}
\usepackage[justification=centering]{caption}
\usepackage{graphicx}
\usepackage{caption}
\usepackage{subcaption}

\usepackage{leftidx}
\usepackage{natbib}
\usepackage[rgb]{xcolor}
\definecolor{MyGreen}{rgb}{0.0,0.6,0.3}
\definecolor{MyPurple}{rgb}{0.6,0,0.3}
\usepackage{fancyref}
\usepackage{hyperref}
\hypersetup{colorlinks=true,citecolor=MyGreen,linkcolor=MyPurple,urlcolor=blue}

\def\beq{\begin{equation}}
\def\eeq{\end{equation}}
\def\ba{\begin{eqnarray}}
\def\ea{\end{eqnarray}}
\def\bal{\begin{align}}
\def\eal{\end{align}}
\def\bxi{{\mbox{\boldmath $\xi$}}}

\begin{document}
\title[Cassini States] {Slowly Rotating Close Binary Stars in Cassini States}
\author[Felce \& Fuller]{
Catherine Felce$^{1}$\thanks{Email: cfelce@caltech.edu} and Jim Fuller$^{1}$
\\$^1$TAPIR, Mailcode 350-17, California Institute of Technology, Pasadena, CA 91125, USA
}
\label{firstpage}
\maketitle
\begin{abstract}

Recent asteroseismic measurements have revealed a small population of stars in close binaries, containing primaries with extremely slow rotation rates. Such stars defy the standard expectation of tidal synchronization in such systems, but they can potentially be explained if they are trapped in a spin-orbit equilibrium known as Cassini state 2 (CS2). This state is maintained by orbital precession due to an outer tertiary star, and it typically results in a very sub-synchronous rotation rate and high degree of spin-orbit misalignment. We examine how CS2 is affected by magnetic braking and different types of tidal dissipation. Magnetic braking results in a slower equilibrium rotation rate, while tidal dissipation via gravity waves can result in a slightly higher rotation rate than predicted by equilibrium tidal theory, and dissipation via inertial waves can result in much slower rotation rates. For seven binary systems with slowly rotating primaries, we predict the location of the outer tertiary predicted by the CS2 theory. In five of these systems, a tertiary companion has already been detected, although it closer than expected in three of these, potentially indicating tidal dissipation via inertial waves. We also identify a few new candidate systems among a population of eclipsing binaries with rotation measurements via spot modulation.

\end{abstract}
\begin{keywords}
stars: evolution --
stars: magnetic field --
stars: kinematics and dynamics --
stars: rotation --
binaries (including multiple): close --
asteroseismology
\end{keywords} 

\section{Introduction}

Stars in close binary systems undergo tidal dissipation, which usually brings the system to a low-energy state characterized by a circular orbit and the synchronization and alignment of each star's spin. Indeed, stars in sufficiently short-period binaries (with orbital periods less than $\sim$3-10 days for main sequence stars) are usually observed to have circular orbits \citep{Duquennoy_1991,Meibom_2005,Meibom_2006,Geller_2009,Milliman_2014,Leiner_2015,Shporer_2016,Windemuth_2019,Justesen_2021}, synchronized rotation rates \citep{Giuricin_1984a,Giuricin_1984b,Claret_1997,Lurie_2017}, and spins aligned with their orbits \citep{Marcussen_2022}. However, there are notable exceptions (discussed below), indicating more complex dynamics are sometimes at play.

The spin and orbital dynamics of stars in triple systems are much more complicated than two-body systems. While the orbital dynamics of such systems has been studied extensively (e.g., \citealt{Naoz_2016}), the spin evolution has received less attention. \cite{Storch_2014} (see also \citealt{Storch_2015,Anderson_2016,Storch_2017}) examined the spin evolution of hot Jupiter host stars during high-eccentricity migration, showing that spin-orbit precession combined with Kozai cycles and tidal circularization can produce a wide range of spin-orbit misalignment angles. 
\cite{Anderson_2017} demonstrated that spin-orbit misalignment can also be created by this process for stellar binaries.

Even for systems with circular orbits, complicated spin-orbit evolution can occur due to the interaction between spin precession of the inner primary and orbital precession of the inner binary induced by the external third body. \cite{Anderson_2018} showed how a resonance between the spin and orbital precession frequencies can excite the obliquity of exoplanet host stars. In a similar manner, spin-orbit misalignment can also be generated due to the precession induced by a protoplanetary disk \citep{Spalding_2014,Spalding_2015,Millholland_2019b,Su_2020,Anderson_2021}.

Cassini states \citep{Colombo_1966,Peale_1969,Ward_1975} are configurations in which the spin-orbit misalignment angle remains constant due to the combined action of spin and orbital precession. In many cases, there not one but \textit{two} stable Cassini states, and systems undergoing any sort of dissipation (e.g., tidal dissipation) will become trapped in one of these states. The first Cassini state typically has nearly synchronous and aligned rotation, similar to the expectation in the absence of the tertiary companion. However, the second Cassini state typically exhibits large spin-orbit misalignment and a sub-synchronous stellar rotation rate. 

In exoplanet systems, planets trapped in Cassini state 2 (CS2) would continue to experience tidal dissipation via obliquity tides, which could contribute to tidal heating and orbital decay of hot Jupiters \citep{Winn_2005,Fabrycky_2007,Millholland_2018} and super-Earth systems \citep{Millholland_2019,Millholland_2020}. \cite{Su_2021} studied this process in great detail, incorporating tidal dissipation torques into the dynamics to derive the modified Cassini states and capture probabilities. There is a substantial parameter space in which a star/planet can be captured into CS2 with large spin-orbit misalignment and slow rotation. 

The same dynamics can occur in close stellar binaries, where stellar spin rates can be measured much more easily. Indeed, recent observations have uncovered a small but growing population of stars in short-period binaries with very slow stellar rotation rates and/or spin-orbit misalignment \citep{Borkovits_2016,Sowicka_2017,Kallinger_2017,Lampens_2018,Fekel_2019,Guo_2019,Li_2020a,Zhang_2020}. These stars not only spin sub-synchronously, but also spin much slower than single stars of similar mass and temperature. We suggest that such systems may in fact be triple systems, with the spin of one (or both) stars in the inner binary trapped in CS2. 

In this paper, we apply the pioneering analysis of \cite{Su_2021} to the case of stellar triples. We extend their work to include the effects of magnetic braking, which can modify or destabilize the Cassini states. We also derive results for a few different tidal dissipation mechanisms, which can similarly affect the spin rates and stability of the Cassini states. We analyze several of the stellar systems discussed above to determine whether they could be trapped in CS2. Several of these systems are already known to have tertiary companions that we show are indeed consistent with the Cassini state hypothesis. In the others, we predict the properties of a putative tertiary star. Finally, we examine the eclipsing binary catalog of \citep{Lurie_2017} to identify a few more candidates for CS2, although we show that many of the sub-synchronous rotators in their sample are false positives.


\section{Cassini Equilibria}
\label{sec:introduction}

\subsection{Systems under Consideration}

We consider binary star systems where the rotation of the primary star can be measured. The semi-major axis of this star around the secondary star is $a_{\rm in}$, the orbital frequency is $n$, the primary's spin magnitude is $\Omega_s$, and its spin direction is $\hat{\boldsymbol{s}}$. The orbital angular momentum of the primary star is $\hat{\boldsymbol{L}}_{\rm in}$, and the angle between the spin and orbital angular momentum vectors is the obliquity $\theta$. The masses of the primary and secondary stars are denoted $M_1$ and $M_2$ respectively, and the radius of the primary star is R. See Fuller \& Felce in prep for a diagram of the spin-orbital configuration.

We then consider a third star, of mass $M_{\rm out}$, orbiting the inner binary system at a semi-major axis of $a_{\rm out}$, with orbital angular momentum $\hat{\boldsymbol{L}}_{\rm out}$. The angle between $\hat{\boldsymbol{L}}_{\rm out}$ and $\hat{\boldsymbol{L}}_{\rm in}$ is denoted by I, the mutual orbital inclination.

\subsection{Full Equations and Defining Parameters of the System}

We follow \cite{Su_2021}, and assume that the spin angular momentum (AM), $S$, is much smaller than the orbital angular momentum, $L$. The spin-orbit evolution equations are given by:
\begin{equation}
\frac{d \hat{\boldsymbol{s}}}{dt}  = \alpha \cos{\theta} \hat{\boldsymbol{s}} \times \hat{\boldsymbol{L}}_{\rm in}  
+ \frac{1}{t_{al}} \hat{\boldsymbol{s}} \times (\hat{\boldsymbol{L}}_{\rm in} \times \hat{\boldsymbol{s}}) 
\label{eq:spin_prec}
\end{equation}

\begin{equation}
\frac{d \hat{\boldsymbol{L}}_{\rm in}}{dt}  = \omega_{lp} \cos{I} \hat{\boldsymbol{L}}_{\rm in} \times \hat{\boldsymbol{L}}_{\rm out}
= -g \hat{\boldsymbol{L}}_{\rm in} \times \hat{\boldsymbol{L}}_{\rm out}
\label{eq:orb_prec}
\end{equation}
where $\theta$ is the obliquity, and the precession rates $\alpha$ and $g$ are defined via:

\begin{equation}
\alpha = \frac{k_2}{2k} \frac{M_2}{M_1} \Bigg(\frac{R}{a_{\rm in}}\Bigg)^3 \Omega_{s}
\end{equation}

\begin{equation}
g = -\frac{3M_{\rm out}}{4(M_1+M_2)} \Bigg(\frac{a_{\rm in}}{a_{\rm out}}\Bigg)^3 \cos I \, \, n 
\textnormal{.}
\label{eq:oms}
\end{equation}
Here, $k = I_s/M_1R^2 $ is the primary's gyration radius, and $k_2$ is the Love number.

Following \cite{Su_2021}, using the weak theory of equilibrium tides, the alignment rate $t_{al}^{-1}$ is related to the spin-up rate $t_s^{-1}$ via
\begin{equation}
    t_{al}^{-1} = t_s^{-1} \Bigg( \frac{2n}{\Omega_s} - \cos{\theta} \Bigg)
\end{equation}
In this case, the spin and orbital elements evolve as:
\begin{equation}
    \frac{d \theta}{dt} =-g \sin{I}\sin{\phi} - \frac{1}{t_s}\sin{\theta} \Bigg( \frac{2 n}{\Omega_s} - \cos{\theta} \Bigg)
    \label{eq:th}
\end{equation}
\begin{equation}
    \frac{d \phi}{dt} = -\alpha \cos{\theta} - g \Big( \cos{I} + \sin{I} \cot{\theta} \cos{\phi} \Big)
    \label{eq:phi}
\end{equation}
\begin{equation}
    \frac{d \Omega_s}{dt} =\frac{\Omega_s}{t_s}
    \Bigg[ \frac{2n}{\Omega_s}\cos{\theta} - (1+\cos^2{\theta}) \Bigg]
    \label{eq:omega}
\end{equation}
where the tidal spin-up time is: 
\begin{equation}
    \frac{1}{t_s} = \frac{3}{4} \frac{k_2}{Qk} \frac{M_2}{M_1} \Bigg( \frac{R}{a_{\rm in}} \Bigg)^3 n
\label{eq:ts}
\end{equation}

The tidal Cassini equilibrium state has $\dot{\Omega}_s = 0$, $\dot{\theta}_s = 0$ and $\dot{\phi}_s = 0$. This requires:

\begin{equation}
    \Omega_s =  \frac{2n\cos{\theta}}{1 + \cos^2{\theta}}
\end{equation}
and therefore $\Omega_s$ approaches zero as $\theta$ approaches $\pi/2$. The other requirement is more complicated and depends on g and $\alpha$. Defining

\begin{align}
    \eta_{sync} &=  -\frac{g}{\alpha} \frac{\Omega_s}{n} \nonumber \\
    &= \frac{3 k}{2 k_2} \frac{M_{\rm out}M_1}{M_2 (M_1+M_2)} \bigg(\frac{a_{\rm in}}{a_{\rm out}}\bigg)^3 \bigg(\frac{a_{\rm in}}{R}\bigg)^3 \cos{I} \textnormal{,}
\label{eq:eta}
\end{align}
\cite{Su_2021} show that the equilibrium of CS2 is 

\begin{equation}
    \cos{\theta_c} \approx \sqrt {\frac{\eta_{sync} \cos{I}}{2}}
\end{equation}

\begin{equation}
    \Omega_{s,c} \approx  2n \cos{\theta_c}
\end{equation}
as long as $\eta_{sync}\ll 1$. Hence, very slow spins can be achieved when $\eta_{sync}\ll1$.

\section{Magnetic Braking}

We now extend our analysis to determine the Cassini equilibria in the presence of both tides and magnetic braking. Magnetic braking slows a star's spin and thus modifies the Cassini equilibria. 

We model this effect by adding a frictional term to the star's spin evolution equation,
\begin{equation}
    \frac{d\Omega_{s,{\rm mag}}}{dt} = - \frac{\Omega_s}{t_{\rm mag}} \, .
\end{equation}
Here, the constant $t_{\rm mag}$ is determined by the angular momentum braking prescription (see, e.g., \citealt{ElBadry_2022}). The maximum (saturated) magnetic braking angular momentum loss rate is (see also \citealt{Sills_2000}):

\begin{equation}
\label{eq:max_braking}
    \dot{J} \simeq -\tau_1 \bigg( \frac{P_{\rm rot}}{1 \, {\rm d}} \bigg)^{-1} \bigg( \frac{P_{\rm crit}}{1 \, {\rm d}} \bigg)^{-2} \bigg( \frac{R}{R_\odot} \bigg)^{1/2} \bigg( \frac{M}{M_\odot} \bigg)^{-1/2} \, .
\end{equation}
with $\tau_1 \sim 10^{35}\, {\rm erg}$ and $P_{\rm crit} \simeq 3 \, {\rm d}$ for Sun-like stars. This leads to a magnetic braking time scale of $t_{\rm mag} \! \sim \! 150$ Myr for Sun-like stars at spin periods of 5 days.

For simplicity, we consider only saturated magnetic braking prescriptions where the spin-down torque is linearly proportional to $\Omega_{\rm s}$. Standard non-saturated braking descriptions have torques proportional to $\Omega_{\rm s}^3$ and could exhibit different Cassini equilibria than those derived below.

Rewriting equation \ref{eq:omega} to include magnetic braking yields
\begin{equation}
    \frac{d\Omega_s}{dt} =\frac{\Omega_s}{t_s}
    \Bigg[ \frac{2n}{\Omega_s}\cos{\theta} - (1+\cos^2{\theta}) - \frac{1}{r_{\rm mag}} \Bigg] \textnormal{,}
    \label{eq:mbo}
\end{equation}
where $r_{\rm mag}$ is the ratio of the magnetic braking and tidal synchronization timescales and is defined as: 
\begin{equation}
   r_{\rm mag} = \frac{t_{\rm mag}}{t_s} \, . 
\end{equation}

\subsection{Condition for Existence of CS1}

There is a minimum value of $r_{\rm mag}$ below which CS1 does not exist. For CS1 to exist, we require that \citep[e.g.,][]{Su_2021}:

\begin{equation}
    \eta < \eta_{crit}
    \textnormal{,}
\end{equation}
 where $\eta = -g/\alpha = (n/\Omega_s) \eta_{sync}$, and $\eta_{crit}$ is defined to be

\begin{equation}
    \eta_{crit} \equiv (\sin^{2/3}{I} + \cos^{2/3}{I} )^{-3/2}
\end{equation}
This implies:

\begin{equation}
    \frac{\Omega_s}{n} > \frac{\eta_{sync}}{\eta_{crit}}
\end{equation}
Solving equation \ref{eq:mbo} for the equilibrium value of $\cos{\theta}$, we get:

\begin{equation}
\label{eq:cosrb}
    \cos{\theta} = \frac{n}{\Omega_s} \pm \sqrt{ \bigg( \frac{n}{\Omega_s} \bigg)^2 - \bigg(1 + \frac{1}{r_{\rm mag}} \bigg) }
\end{equation}
For this to be real, we need
\begin{equation}
    r_{\rm mag} > \Bigg[ \bigg(\frac{n}{\Omega_s} \bigg)^2 - 1 \Bigg]^{-1} \,
\end{equation}
giving a lower bound on $r_{\rm mag}$ for CS1 to exist:

\begin{equation}
    r_{{\rm mag},{\rm min}} = \frac{\eta_{sync}^2}{\eta_{crit}^2 - \eta_{sync}^2}
    \approx \frac{\eta_{sync}^2}{\eta_{crit}^2}
    \textnormal{,}
    \label{eq:r_Bmin}
\end{equation}
where in the last approximation we have assumed $\eta_{sync} \ll 1$.

\subsection{Modified Cassini Equilibrium}

Solving equation \ref{eq:mbo} for equilibrium gives:
\begin{equation}
\label{eq:Omseq}
    \frac{\Omega_{s,{\rm eq}}}{n} = \Bigg[ \frac{1 + \cos^2{\theta_{\rm eq}} + r_{\rm mag}^{-1}}{2\cos{\theta_{\rm eq}}} \Bigg]^{-1} 
\end{equation}
Following Appendix A of \citep{Su_2021}, we consider slow tidal dissipation such that $(|g|t_s)^{-1} \ll 1$, such that CS2 has $\phi \simeq \pi$. In the limit of $\eta \ll 1$, $\theta \simeq \pi/2 - \eta \cos{I}$. Then solving equation \ref{eq:mbo} for $\Omega_{s,{\rm eq}}$ again gives

\begin{equation}
    \frac{2n}{\Omega_{s,{\rm eq}}} \frac{\eta_{sync}n}{\Omega_{s,{\rm eq}}}\cos{I} 
    - 1 - \frac{1}{r_{\rm mag}}  \simeq 0
    \textnormal{,}
\end{equation}
and therefore:

\begin{equation}
\label{eq:omseqmb}
    \frac{\Omega_{s,{\rm eq}}}{n} \simeq \sqrt{\frac{2\eta_{sync} \cos{I}}{1 + r_{\rm mag}^{-1}}}
\end{equation}
In the presence of strong magnetic braking, the CS2 rotation rate is thus decreased by a factor of $\sqrt{t_{\rm mag}/t_s}$. This will be especially important in more widely separated systems where $t_s$ is larger. 

The CS2 equilibrium obliquity angle is still given by $\cos \theta_{\rm eq} \simeq \eta \cos I$, and hence
\begin{equation}
    \cos \theta_{\rm eq} \approx \sqrt {\frac{\eta_{sync} \cos{I} (1 + r_{\rm mag})^{-1}}{2}} \, .
\end{equation}
Stronger magnetic braking therefore corresponds to a smaller equilibrium obliquity. This can be understood because a smaller equilibrium rotation rate requires a smaller spin-orbit misalignment angle in order to maintain enough spin-orbit precession to maintain the equilibrium.

\subsection{Numerical Solutions with Magnetic Braking}

\begin{figure}
    \centering
    \begin{subfigure}[b]{0.5\textwidth}
        \hspace*{-0.4cm}
        \includegraphics[scale=0.27]{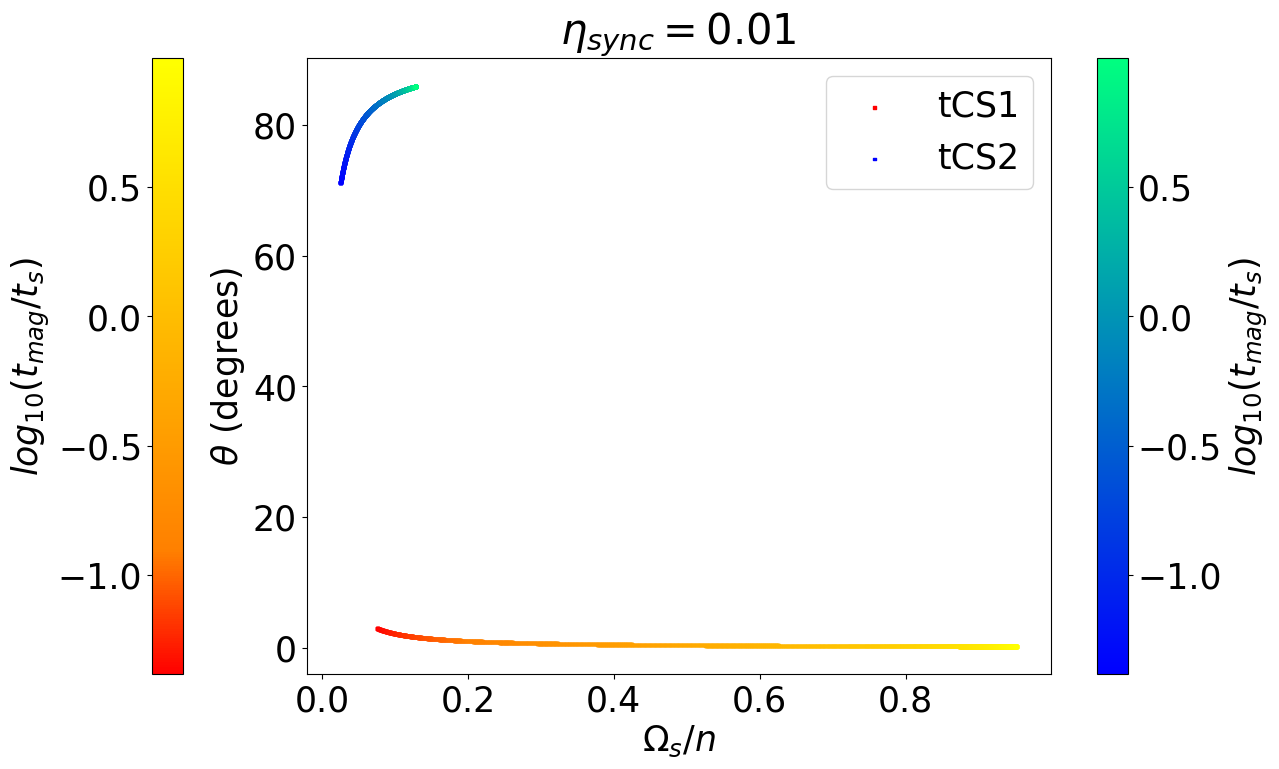}
    \end{subfigure}
    \begin{subfigure}[b]{0.5\textwidth}
        \hspace*{-0.4cm}
        \includegraphics[scale=0.27]{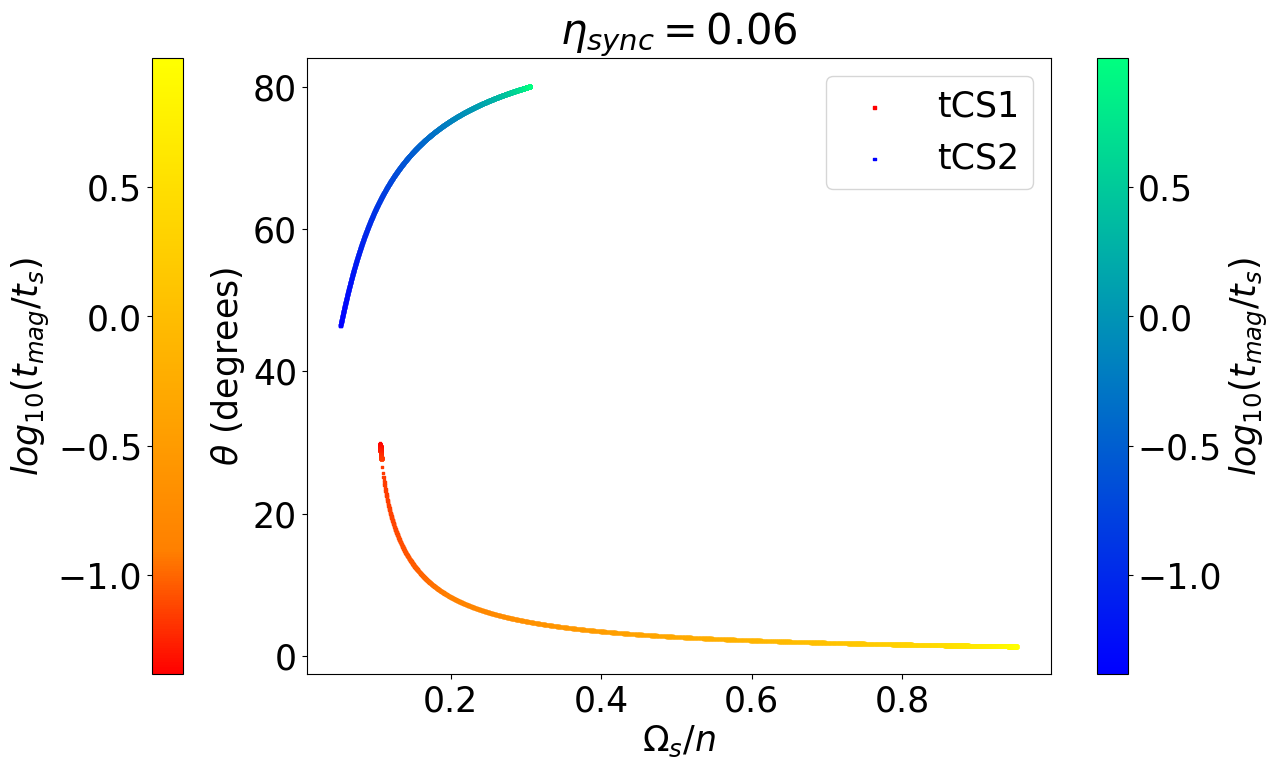}
    \end{subfigure}
    \begin{subfigure}[b]{0.5\textwidth}
        \hspace*{-0.4cm}
        \includegraphics[scale=0.27]{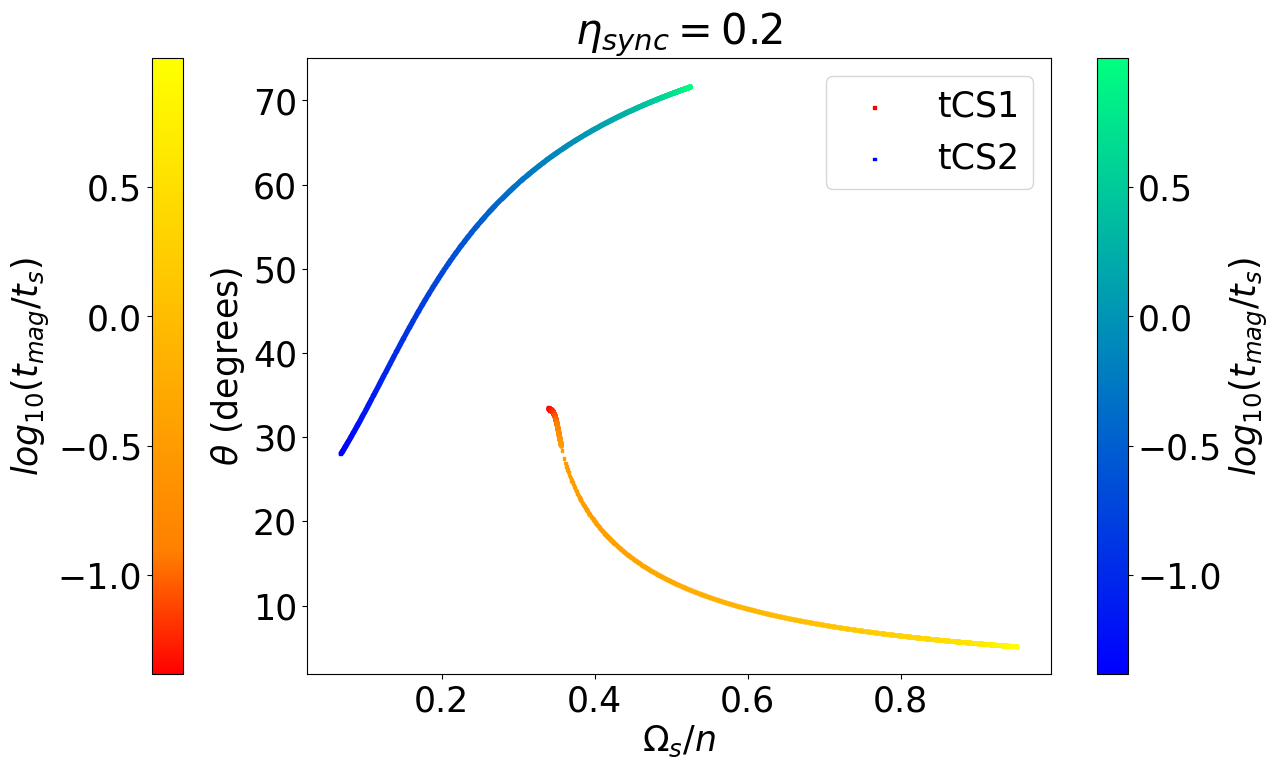}
    \end{subfigure}
    \caption{Phase-space position of tidal Cassini State equilibria: CS1 in red/orange and CS2 in blue/green, where the color scale indicates different values of $r_{\rm B} \equiv t_{\rm mag}/t_{\rm s}$. Stronger magnetic braking moves the Cassini states to slower rotation and larger/smaller misalignment angles for CS1 and CS2, respectively. We take $I=20^{\circ}$, and the values of $\eta_{\rm sync}$ are 0.01, 0.06 and 0.2 in the top, middle and bottom panels respectively.}
    \label{fig:tCSs_rB}
\end{figure}

Magnetic braking has the effect of moving the equilibrium $\Omega_s$ to lower values. The position of CS1 and CS2 as a function of $r_{\rm mag}$, for different values of $\eta_{\rm sync}$, is shown in Figure \ref{fig:tCSs_rB}. These are calculated by numerically solving for the stationary point of $\Omega_s$ using equation \ref{eq:mbo}, in combination with stationarity of $\phi$ from equation \ref{eq:phi}:

\begin{equation}
    \sin{\theta_{\rm eq}}\cos{\theta_{\rm eq}} - \frac{\eta_{\rm sync}n}{\Omega_{s,{\rm eq}}}\sin{(\theta_{\rm eq} \pm I)} = 0
    \label{eq:CS}
    \textnormal{.}
\end{equation}
Here, the $\pm$ sign refers to CS1 and CS2, which have $\phi \simeq 0$ and $\phi \simeq \pi$, respectively.

Cassini state equilibria with varying $r_{\rm mag}$ are shown in Figure \ref{fig:tCSs_rB}. For these calculations we used $I = 20^{\circ}$ and $\eta_{\rm sync}=0.01, 0.06$ and $0.2$ for the top, middle and bottom panels respectively. Both CS1 and CS2 are moved to smaller spin rotation rates as the strength of magnetic braking increases (decreasing $r_{\rm mag} \equiv t_{\rm mag}/t_{\rm s}$). However, for CS1, increased magnetic braking increases the spin-orbit obliquity angle, whereas for CS2 this relationship is reversed. We have a lower bound on $r_{B}$ from equation \ref{eq:r_Bmin}, below which CS1 no longer exists, e.g. $r_{B} \geq 0.011$ for the $\eta_{sync}=0.06$ case.

\begin{figure}
     \centering
     \begin{subfigure}[b]{0.5\textwidth}
         \centering
         \includegraphics[width=\textwidth]{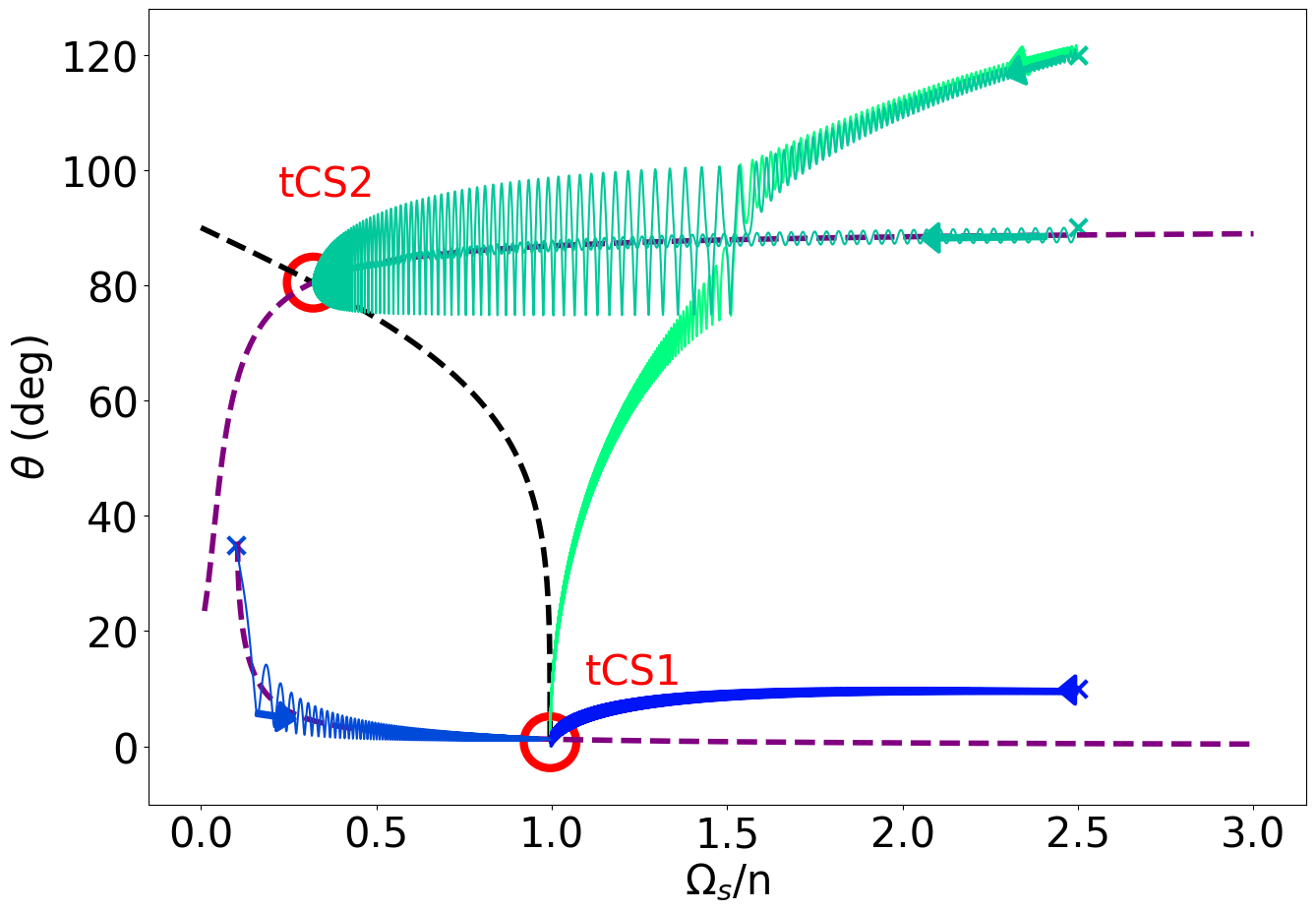}
         \label{fig:y equals x}
     \end{subfigure}
     \hfill
     \begin{subfigure}[b]{0.5\textwidth}
         \centering
         \includegraphics[width=\textwidth]{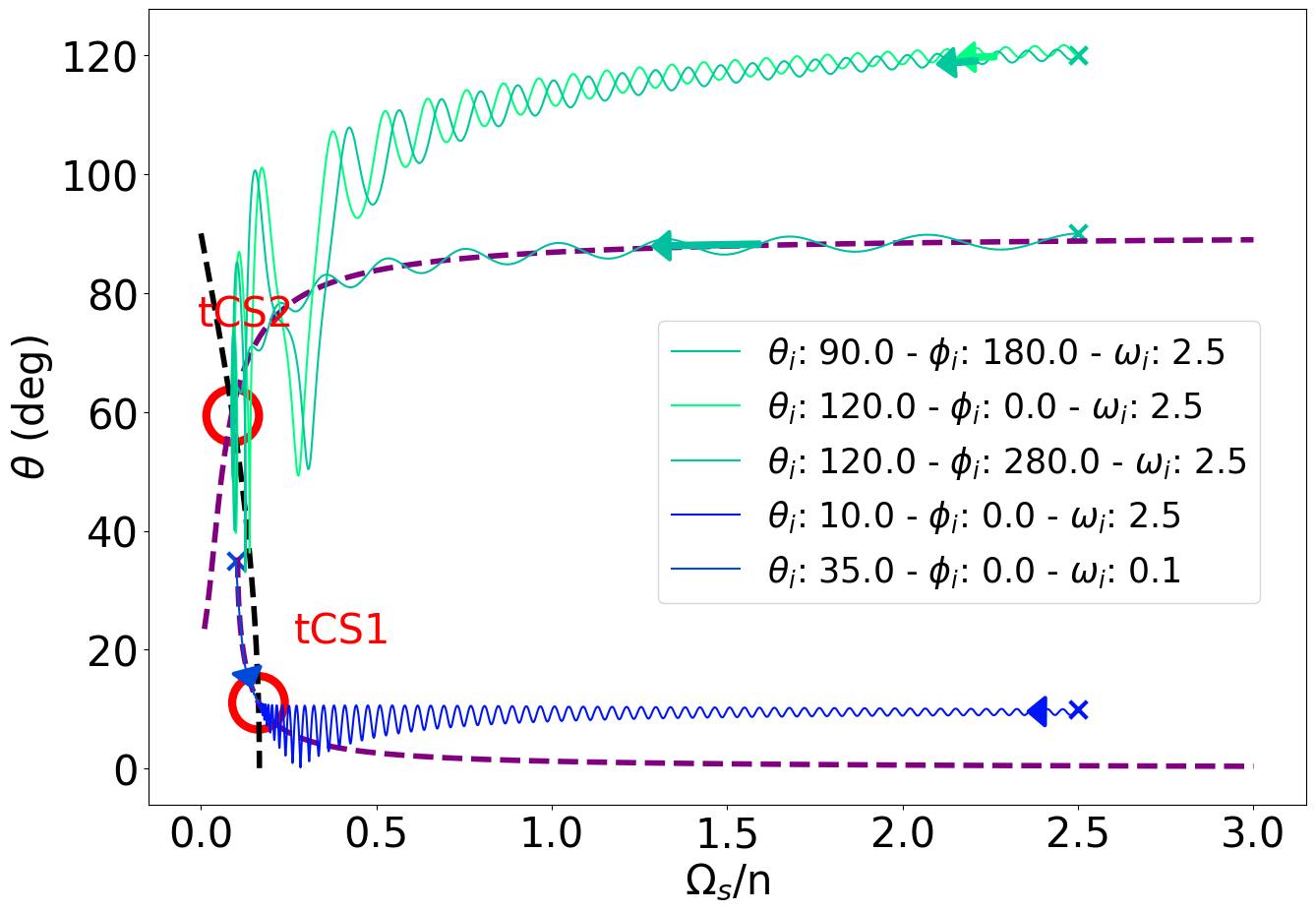}
         \label{fig:five over x}
     \end{subfigure}
        \caption{
        Phase-space trajectories for different values of the ratio $t_{\rm mag}/t_s \equiv r_{\rm mag}$. The top panel has $r_{\rm mag} = 100$ and the bottom panel has $r_{\rm mag} = 0.1$. Initial conditions for each trajectory are shown (obliquity: $\theta_i$, phase: $\phi_i$ and rotation rate: $\omega_i=\Omega_s/n$). We have set $\eta_{\rm sync} = 0.06$, $I=20^\circ$, $|g|t_{\rm s}=100$. The black dashed line shows points where $\Omega_s$ is stationary; the purple dashed lines show where $\theta$ is stationary. The red circles show the equilibrium points, tidal Cassini States 1 and 2. Arrows indicate the direction of evolution for each trajectory, which move towards one of the Cassini states.
        }
        \label{fig:mag_traj}
\end{figure}

We also investigate the time-dependent spin-orbit evolution of a few systems, as shown in Figure \ref{fig:mag_traj}. To make Figure \ref{fig:mag_traj}, we integrated equation \ref{eq:mbo} along with equations \ref{eq:th} and \ref{eq:phi}, using $|g|t_s=100$, $I = 20^{\circ}$, $\eta_{\rm sync}=0.06$ and various values of $r_{\rm mag}$. The different colored trajectories correspond to different initial conditions. Some trajectories are captured into CS2 and others go to CS1. Both stationary points are at slower spins due to magnetic braking.

\section{Dynamical tides}

The weak tidal friction theory used above and in nearly all prior work assumes a tidal lag time that is the same for each component of the tidal forcing. This must be modified when dissipation occurs via dynamical tides, where the tidal dissipation is strongly dependent on the tidal forcing frequency. The relative synchronization and alignment torques are modified, along with the Cassini equilibria. Here we derive the modified Cassini equilibria for dissipation via internal gravity waves and inertial waves.

\subsection{Gravity waves}

To investigate possible implications of dynamical tides for tidal Cassini states, we begin with Equation 27 of \cite{Lai_2012}
\begin{align}
\label{eq:torque}
   \frac{I}{T_0} \frac{d \Omega_s}{dt} &= (1+c)^4 (n - \Omega_s) \tau_{22} + s^2 (1+c)^2 (2n - \Omega_s) \tau_{12} \nonumber \\
   & - (1-c)^4 (n + \Omega_s) \tau_{-22} - s^2 (1-c)^2 (2n + \Omega_s) \tau_{-12} \nonumber \\
   & - 4 \Omega_s (s^4 \tau_{20} + s^2 c^2 \tau_{10}) \, ,
\end{align}
where $T_0 = (3 \pi G/20)(M_2^2 R^5/a_{\rm in}^5)$, $s= \sin \theta$, and $c = \cos \theta$. Here $\tau_{m m'}$ is the effective tidal lag time of each tidal component, whose frequency in the rotating frame of the star is $\omega_{m m'} = m' n - m \Omega_s$.
We are interested in CS2 where we expect $\cos \theta \ll 1$ and $\Omega_s \ll n$. In this case, the tidal forcing frequency is nearly equal for $\pm m$ components, i.e., $\omega_{m m'} \simeq \omega_{-m m'} \simeq m' n$. We therefore assume $\tau_{22} \simeq \tau_{-22} \simeq \tau_{12} \simeq \tau_{-12}$. Expanding to first order in $\cos \theta$ and $\Omega_s$, equation \ref{eq:torque} becomes
\begin{equation}
    \frac{I}{T_0} \frac{d \Omega_s}{dt} \simeq 16 c n \tau_{22} - 4 \Omega_s \tau_{22} - 4 \Omega_s \tau_{20} - 4 c^2 \Omega_s \tau_{10} \, .
\end{equation}

We now consider different tidal dissipation mechanisms. For tidally excited gravity waves, the dissipation is much more efficient at high frequencies \citep{Goodman_1998}, so we expect $\tau_{22} \gg \tau_{20},\tau_{10}$. In this case, the spin-equilibrium occurs when 
\begin{equation}
    \Omega_s \simeq 4 n \cos \theta
\end{equation}
Then at CS2 where
\begin{equation}
\label{eq:tcs2}
    \cos \theta \simeq n \eta_{\rm sync} \cos I/\Omega_s
\end{equation}
we have 
\begin{equation}
    \cos \theta \simeq \sqrt{\eta_{\rm sync} \cos I/4} \, ,
\end{equation}
\begin{equation}
    \Omega_s \simeq n \sqrt{4 \eta_{\rm sync} \cos I} \, .
\end{equation}
The spin rate is thus a factor of $\sqrt{2}$ larger than the weak tidal friction case.

\subsection{Inertial waves}

Here we follow the procedure of Fuller \& Felce (in prep) to examine the case of inertial wave dissipation. Following \cite{Lai_2012}, we consider  the dissipation via inertial waves due to the $m=1$, $m'=0$ component of the tidal forcing, which is the only component that resonates with inertial waves when $\Omega_{\rm s} \ll n$. Considering equilibrium tidal dissipation of tidal lag time $\tau$, in addition to extra tidal dissipation via inertial waves via a factor $X_{10} = \tau_{10}/\tau - 1$, the spin evolution equations are modified to 

\begin{equation}
\label{eq:omsdot10}
    \frac{d\Omega_s}{dt} = \frac{\Omega_s}{t_s} \bigg[ \frac{2 n}{\Omega_s} \cos \theta - (1 + \cos^2 \theta) -\frac{X_{10}}{2} \sin^2 \theta \cos^2 \theta \bigg] \, ,
\end{equation}
\begin{equation}
\label{eq:thetadot10}
    \frac{d \theta}{dt} = -g \sin I \sin \phi - \frac{\sin \theta}{t_s} \bigg[ \frac{2 n}{\Omega_s} - \cos \theta + \frac{X_{10}}{2} \cos^3 \theta \bigg] \, ,
\end{equation}

\begin{figure}
     \centering
     \begin{subfigure}[b]{0.45\textwidth}
         \centering
         \includegraphics[width=\textwidth]{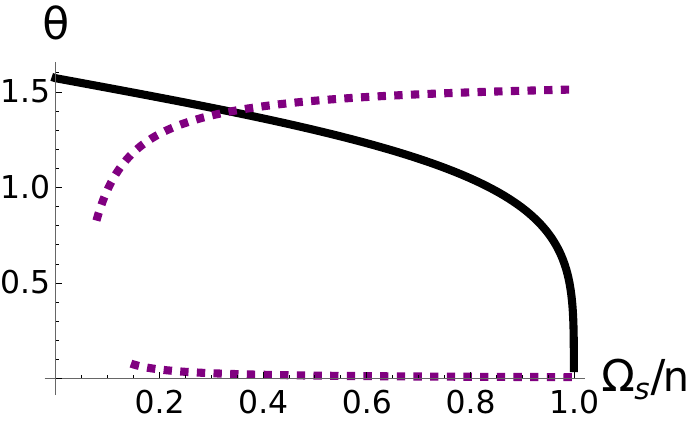}
         \label{fig:friction_0}
     \end{subfigure}
     \hfill
     \begin{subfigure}[b]{0.45\textwidth}
         \centering         \includegraphics[width=\textwidth]{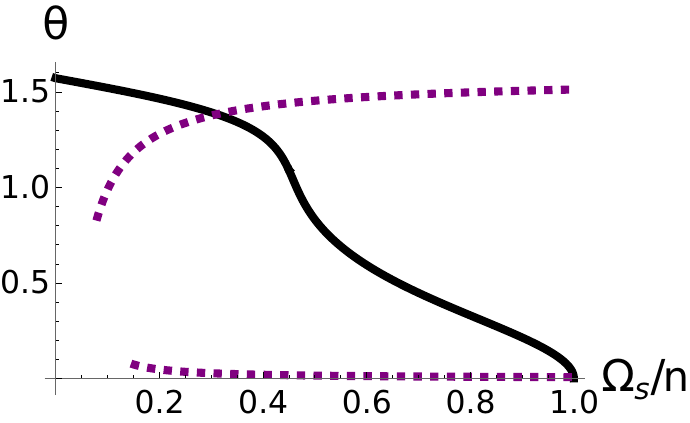}
         \label{fig:friction_10}
     \end{subfigure}
     \hfill
     \begin{subfigure}[b]{0.45\textwidth}
         \centering
         \includegraphics[width=\textwidth]{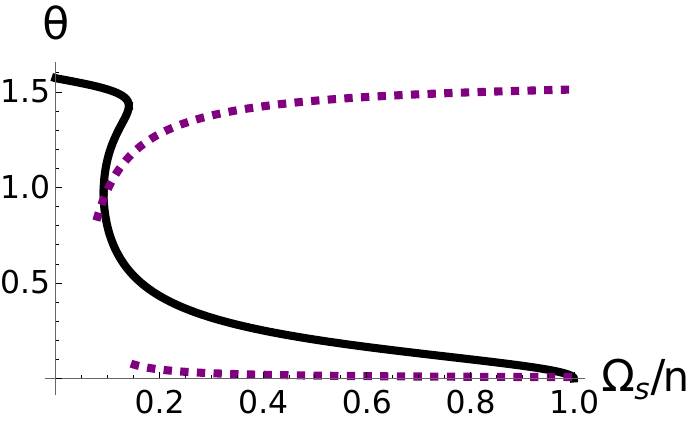}
         \label{fig:friction_100}
     \end{subfigure}
        \caption{Spin rotation equilibria in a phase-space of obliquity, $\theta$, vs rotation frequency. The thick black lines show the spin equilibrium from equation \ref{eqn:omseq10} for $\eta_{\rm sync}=0.06$ and $I=20^\circ$ and different values of the efficiency of inertial wave dissipation, parameterized by $X_{10}$. The values for $X_{10}$ are: top: $X_{10}=0$, middle: $X_{10}=10$, bottom: $X_{10}=100$. The purple dashed lines show where $\theta$ is stationary. Note that CS2 (where the upper purple and black lines intersect) moves to smaller spin rate with larger $X_{10}$.}
        \label{fig:tau_omega}
\end{figure}

The equilibrium spin rate is modified to
\begin{equation}
\label{eqn:omseq10}
    \Omega_{s,{\rm eq}} = \frac{4 \cos \theta \, n}{2 + 2 \cos^2 \theta + X_{10} \sin^2 \theta \cos^2 \theta } \, .
\end{equation}
As in the case of magnetic braking, this modification changes the shape of the stationary $\Omega_s$ line, as shown in Figure \ref{fig:tau_omega}. At high values of $X_{10}$, the stationary omega line becomes multi-valued in $\theta$ and may cross CS1 or CS2 lines twice, allowing for multiple Cassini equilibria. 

We can analytically solve for $\Omega_{s,{\rm eq}}$ assuming $\cos \theta_{\rm eq} \ll 1$ at CS2. Following Appendix A of \cite{Su_2020}, we set $(|g|t_s)^{-1} = 0$, and expand equation \ref{eq:phi} in powers of small $\eta = \eta_{sync}/\Omega$, to get the CS2 equilibrium of $\phi=\pi$, $\theta= \pi/2 - \eta \cos{I}$. Then using $\cos(\pi/2 - \eta \cos I) \simeq \sin (\eta \cos I)$ and rearranging equation \ref{eqn:omseq10} gives
\begin{multline}
    \frac{4n}{\Omega_{s,{\rm eq}}} \sin{(\eta \cos{I})} - 2 ( 1 + \sin^2{(\eta \cos{I})})  - \\  [ \cos{(\eta \cos{I})} \sin{(\eta \cos{I})} ]^2 X_{10} = 0
    \textnormal{,}
\end{multline}
\begin{figure}
    \hspace{-0.25in}
    \includegraphics[scale=0.3]{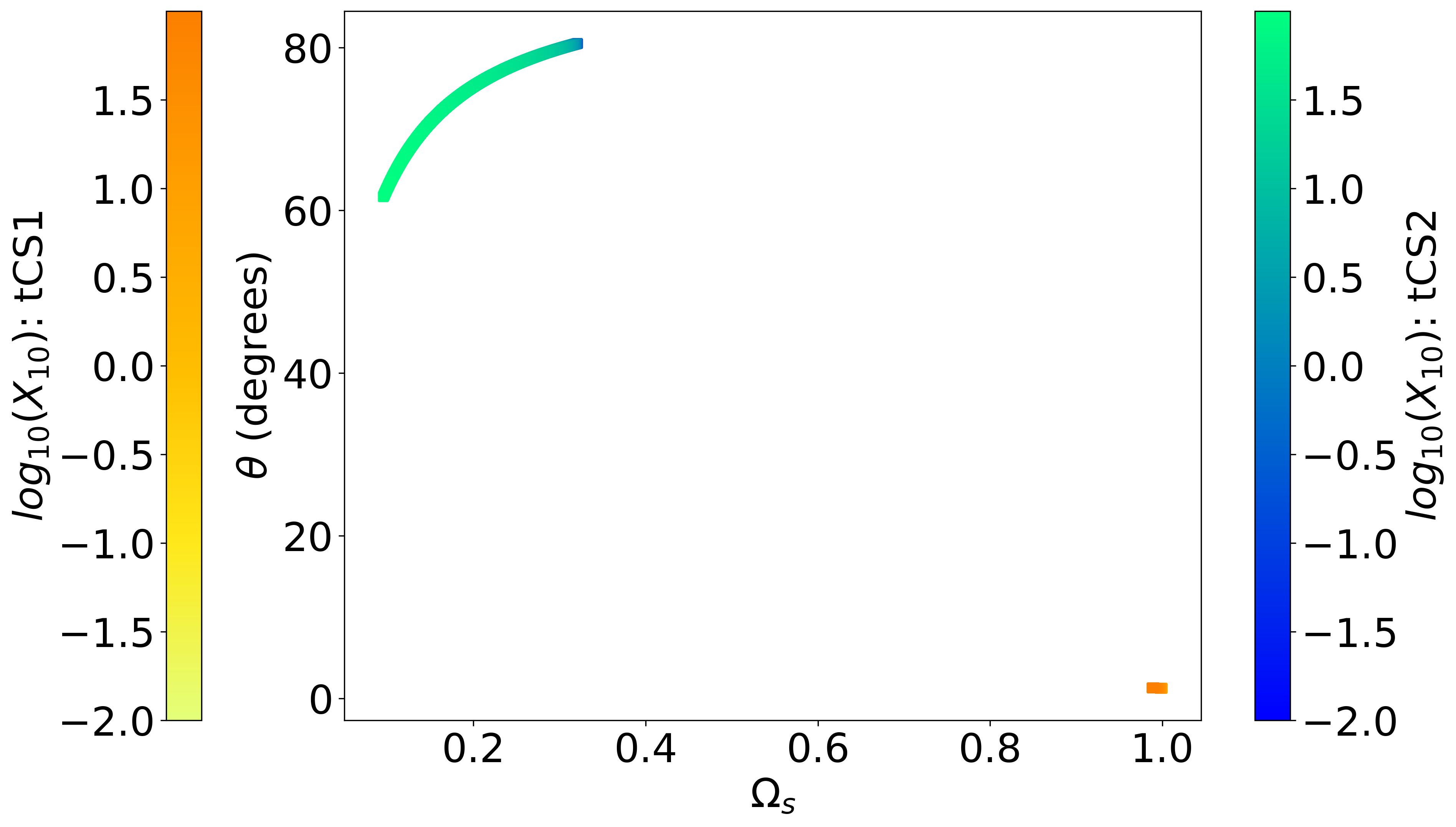}
    \caption{Equilibrium spin rate and misalignment angle for Cassini state 1 (reds) and 2 (blues) including dissipation via inertial waves. The color-scale indicates the value of $X_{10} = \tau_{10}/\tau \boldsymbol{-1}$, which characterizes the strength of the inertial wave tidal friction. We use $\eta_{sync} = 0.06$, and $I = 20^\circ$.}
    \label{fig:friction_tCEs}
\end{figure}
Now using $\eta = \eta_{sync} n/\Omega_s$ 
to second order in $\eta_{sync}$ we get:

\begin{equation}
    \frac{4n^2 \eta_{sync}} {\Omega_{s,{\rm eq}}^2} \cos{I} 
    - 2 - \Bigg( \frac{\eta_{sync} \cos{I}}{\Omega_{s,{\rm eq}}} \Bigg)^2 X_{10} = 0
    \textnormal{.}
\end{equation}
Rearranging for $\Omega_{s,{\rm eq}}$ gives:
\begin{equation}
    \frac{\Omega_{s,{\rm eq}}^2}{n^2} = 2 \eta_{sync} \cos{I} - \frac{X_{10}}{2} (\eta_{sync}\cos{I})^2
    \textnormal{.}
\end{equation}
and a corresponding misalignment angle
\begin{equation}
    \theta_{\rm eq} = \pi/2 -  \frac{\eta_{sync} \cos{I}}{\sqrt{2 \eta_{sync} \cos{I} - X_{10} (\eta_{sync}\cos{I})^2/2}}
\end{equation}

We see that when $X_{10} \eta_{\rm sync} \cos I/4 > 1 $, the CS2 equilibrium no longer exists, or at least it does not satisfy the assumption that $\cos \theta_{\rm eq} \ll 1$. In this case, there can be a different equilibrium as discussed in Fuller \& Felce, in prep. Figure \ref{fig:tau_omega} demonstrates this effect: at moderate values of $X_{10}$ the CS2 equilibrium remains near $\theta_{\rm eq} \approx \pi/2$, but the changing shape of the stationary $\Omega_s$ line causes $\theta_{\rm eq}$ to decrease, eventually approaching $\theta_{\rm eq} \approx I$ or $\theta_{\rm eq} \approx \pi - I$ in the limit of very large $X_{10}$.

We compute the CS equilibria for inertial wave dissipation for different values of $X_{10}$. As before, we assume a large $t_s$ and take $\phi=0, \pi$. Then we numerically solve equations \ref{eq:phi} and \ref{eq:omsdot10} in equilibrium. Figure \ref{fig:friction_tCEs} shows results using $\eta_{sync} = 0.06$, and $I = 20^\circ$. As $X_{10}$ increases, the location of CS1 and CS2 both move towards lower spin rates. The misalignment angle of CS1 increases, while it decreases for CS2. 

Finally, including inertial wave dissipation will alter the  minimum value of $t_{s,c}$ below which CS2 is destabilized. As in \cite{Su_2021}, this limit is obtained from equation \ref{eq:th}, which for equilibrium tides yields
\begin{equation}
    \label{eq:tsc}
    t_{s,c} = \frac{\sin \theta}{g \sin I} \bigg( \frac{2 n}{\Omega_s} - \cos \theta \bigg) \, .
\end{equation}
Including inertial waves and using equation \ref{eq:thetadot10}, this is modified to
\begin{equation}
    \label{eq:tsci}
    t_{s,c} = \frac{\sin \theta}{g \sin I} \bigg( \frac{2 n}{\Omega_s} - \cos \theta + \frac{X_{10}}{2} \cos^3 \theta \bigg) \, .
\end{equation}
When $X_{10} \eta_{\rm sync} \cos I /4 \ll 1$, one can show that the new term in equation \ref{eq:tsci} is smaller than the second and can be ignored. In the limit $X_{10} \eta_{\rm sync} \cos I /4 \gg 1$, the equilibrium spin rate from Fuller \& Felce in prep is $\Omega_s \simeq 4 n/(X_{10} \sin^2 \theta \cos \theta)$. Using this relation yields 
\begin{equation}
    \label{eq:tsci2}
    t_{s,c} \simeq \frac{\sin \theta}{g \sin I} \bigg( \frac{2 n}{\Omega_s \sin^2 \theta} - \cos \theta \bigg) \, .
\end{equation}
Hence the critical value of $t_{s,c}$ is only slightly altered for systems in CS2 where $\sin \theta$ is of order unity.

\section{Comparison with Observations}

In the following section we examine several sub-synchronously rotating stars we have discovered in the literature. Additional such systems may also exist, so this list may be very incomplete. Based on the properties of each system we examine, we derive constraints for the mass and inclination of a putative third body, assuming that the systems are trapped in CS2. Since all of the systems we consider have intermediate-mass primary stars, we expect magnetic braking to have a small effect. This justifies the assumption that sub-synchronicity in these systems is better explained by their being in CS2 rather than in a magnetically-braked CS1.

\subsection{System Properties}
\label{sec:meas}

\begin{table*}
\caption{Parameter values for 6 systems with very slowly rotating stars in close binaries. The outer orbital period, $P_{\rm out}$, is included for binaries that are part of known triple systems.
Reference labels are 1: \citealt{Lampens_2018}, 2: \citealt{Gaia_DR2}, 3: \citealt{Li_2020a}, 4: \citealt{Sowicka_2017}, 5: \citealt{Guo_2019}, 6: \citealt{Zhang_2019}, 7: \citealt{Borkovits_2016}, 8: \citealt{Windemuth_2019}, 9: \citealt{Kallinger_2017}, 10: \citealt{Zhang_2020}, 11: \citealt{Fekel_2019}, 12: \citealt{Sekaran_2020}.  Distances are taken from \citealt{Gaia_EDR3}}. 
\hskip-0.6cm \begin{tabular}{@{}ccccccccc@{}}
\toprule
$\textbf{Name }$ & $\mathbf{M_1}$ & $\textbf{M}_{\textbf{2}}$ & $\textbf{R}$& $\textbf{P}_{\textbf{\rm orb}}$ & $\textbf{P}_{\textbf{\rm rot}}$ & $\textbf{P}_{\textbf{\rm out}}$ & $\textbf{distance}$ & $\textbf{Refs}$\\
& $(\textbf{M}_{\odot})$ & $(\textbf{M}_{\odot})$ & $(\textbf{R}_{\odot})$ & $(\textbf{d})$  & $(\textbf{d})$  & $(\textbf{d})$  & $(\textbf{pc})$  & \\
\midrule
KIC 4480321& $1.5^{+0.3}_{-0.2} $& $1.5^{+0.3}_{-0.2} $& $1.9^{+0.5}_{-0.5} $& $9.166^{+6e-05}_{-6e-05} $& $121.0^{+4.0}_{-4.0} $& $2280.0^{+29.0}_{-29.0} $& $700^{+11}_{-11}$ & 1, 2, 3\\
KIC 8197761& $1.384^{+0.281}_{-0.276} $& $0.28^{+7e-01}_{-0e+00} $& $1.717^{+0.858}_{-0.41} $& $9.869^{+3e-07}_{-3e-07} $& $301.0^{+3.0}_{-3.0} $ & -& $399^{+3}_{-3}$ & 4\\
KIC 4142768& $2.05^{+0.03}_{-0.03} $& $2.05^{+0.03}_{-0.03} $& $2.96^{+0.04}_{-0.04} $& $13.996^{+6e-05}_{-6e-05} $& $2702.7^{+1300.0}_{-662.0} $ &  -& $1430^{+24}_{-24}$ & 3, 5\\
KIC 8429450& $1.68^{+0.2}_{-0.13} $& $1.462^{+0.174}_{-0.113} $& $2.438^{+0.083}_{-0.081} $& $2.705^{+2e-07}_{-2e-07} $& $38.0^{+128.0}_{-17.0} $& $3088.0^{+1700.0}_{-1700.0} $& $1520^{+30}_{-30}$ & 3, 6, 7, 8\\
HD 201433& $3.05^{+0.025}_{-0.025} $& $0.7^{+0.3}_{-0.3} $& $2.6^{+0.2}_{-0.2} $& $3.313^{+5e-04}_{-5e-04} $& $292.0^{+76.0}_{-76.0} $& $154.2^{+0.03}_{-0.03} $& $122^{+2}_{-2}$ & 9\\
KIC 9850387& $1.66^{+0.01}_{-0.01} $& $1.062^{+0.003}_{-0.005} $& $2.154^{+0.002}_{-0.004} $& $2.749^{+5e-06}_{-5e-06} $& $188.68^{+74.5}_{-41.6} $& $671.0^{+2.0}_{-2.0} $& $1870^{+50}_{-50}$ & 10, 3, 7, 12\\
HD 126516 & $1.34^{+0.2}_{-0.2} $& $0.28^{+0.03}_{-0.03} $& $1.66^{+0.08}_{-0.08} $& $2.124^{+1e-07}_{-1e-07} $& $18.3^{+2.8}_{-7.7} $& $702.71^{+0.25}_{-0.25} $& $102^{+2}_{-2}$ & 11\\
\bottomrule
\end{tabular}
\label{tab:system_params}
\end{table*}

The system properties described in this section are summarized in Table \ref{tab:system_params}. In what follows, $M_1$ is the mass of the brighter star in the inner binary, $P_{\rm rot}$ is its rotation period, $P_{\rm orb}$ is the orbital period of the inner binary, and $P_{\rm out}$ is the outer orbital period.

All systems have intermediate-mass ($1.4-3 \, M_\odot$) primaries on the main sequence, with inner orbital periods in the range $2-14$ days, most of which are nearly circular. Five of the systems have known tertiary companions with orbits of $0.4-9$ years, but the others may also have nearby tertiaries that have not been detected. Nearly all of these systems have pulsating primaries ($\gamma$-Doradus or SPB pulsations), and the very slow rotation period was measured via asteroseismology. Below we describe key details of each system.


\subsubsection{KIC 4480321}
KIC 4480321 forms part of a known triple system, with the tertiary A-type star more luminous than either star in the inner binary \citep{Lampens_2018}, which is composed of nearly equal-mass F-type stars whose temperatures lie in the range $6300 \, {\rm K} \lesssim T_{\rm eff} \lesssim 6900 \, {\rm K}$. Unfortunately, this system is not eclipsing, and the outer star would contaminate a Gaia radius estimate, so the inner binary masses and radii are very uncertain. Based on stellar models in \cite{Fuller_2017}, we estimate masses in the range $1.3-1.8 \, M_\odot$ and radii in the range $1.4-2.4 \, R_\odot$, which are consistent with the observed temperatures and orbital velocities.

\cite{Lampens_2018} also gives a constraint on the inclination relative to the line of sight, $i$, of the system as $36^{\circ} < i_{\rm in} < 63^{\circ}$, and $42^{\circ} < i_{\rm out} < 57^{\circ}$.
One of the inner binary stars is a $\gamma$-Doradus pulsator whose spin period was asteroseismically measured to be $P_{\rm rot}=121^{+4.0}_{-4.0} \, {\rm d}$ by \cite{Li_2020a}.

\subsubsection{KIC 8197761}
KIC 8197761 is a $\gamma$ Doradus - $\delta$ Scuti hybrid pulsator in a binary system.
We use  the stellar masses, radius, orbital and spin periods measured by \cite{Sowicka_2017}, who used RVs in combination with transit photometry. They also asteroseismically measured a rotation period of the primary to be $P_{\rm rot} = 301.0^{+3.0}_{-3.0} \, {\rm d}$. There are no published limits on a possible third body in this system.

\subsubsection{KIC 4142768}
Unlike the other systems studied here, KIC 4142768 consists of two evolved A-type stars in an eccentric orbit with $e=0.582 \pm 0.002$. In addition to $\gamma$-Doradus pulsations, KIC 4142768 is a heartbeat star exhibiting tidally excited pulsations \citep{Guo_2019}. Masses and radii were measured via RVs and light curve modeling, and the asteroseismically measured spin period \citep{Li_2020a} is extremely slow, $P_{\rm rot} = 2702.7^{+1300.0}_{-662.0} \, {\rm d}$, though with large uncertainty.

\subsubsection{KIC 8429450}
KIC 8429450 is a $\gamma$-Doradus star in an eclipsing binary which is part of a triple system. We adopt the spectroscopically measured primary mass \citep{Zhang_2019}, and 
the orbital period  from \cite{Windemuth_2019}, who also measure a mass ratio $q = 0.87$. These properties are derived via photometric modeling and the observed spectral energy distribution of the system, combined with stellar evolution models.
The rotation period is asteroseismically measured by \cite{Li_2020a}: $P_{\rm rot}=38^{+128}_{-17} \, {\rm d}$. The orbital period of the outer tertiary is roughly constrained by \cite{Borkovits_2016} to be $P_{\rm out}=3088^{+1700}_{-1700} \, {\rm d}$ based on eclipse timing measurements, the longest measured period in our sample. The projected motion of the inner binary along the line of sight, $a_{12} \sin{i}$, is measured to be $128^{+75}_{-75} {\rm R}_{\odot}$.

\subsubsection{HD 201433}
HD 201433 is the primary in a triple system containing a slowly pulsating B9V star, investigated in great detail by \cite{Kallinger_2017}. The primary mass of $M_1=3.05^{+0.025}_{-0.025}$ is the highest in our sample and the asteroseismically measured rotation period is $P_{\rm rot}=292^{+76}_{-76} \, {\rm d}$. This system also contains the shortest period tertiary in our sample, with $P_{\rm out}=154.2^{+0.03}_{-0.03} \, {\rm d}$.

\subsubsection{KIC 9850387}
KIC 9850387 is a $\gamma$ Doradus-$\delta$ Scuti star in an eclipsing binary \citep{Zhang_2020}. Based on the gravity mode period spacing, the primary could be on the pre-main sequence with an estimated age of only $\approx 7$ Myr. However, this conclusion is not confirmed by the detailed binary and asteroseismic modeling of \cite{Sekaran_2020,sekaran:21}, who accurately measure the masses and radii of the stars in the inner binary. 
The rotation period is estimated by \cite{Li_2020a} to be $P_{\rm rot}=188.68^{+74.5}_{-41.6} \, {\rm d}$. The outer orbital period of the triple system is derived by \cite{Borkovits_2016} using eclipse timing variations to be $P_{\rm out}=671.0^{+2.0}_{-2.0} \, {\rm d}$, with $a_{12} \sin{i} = 98^{+1}_{-1} \, R_\odot $.

\subsubsection{HD 126516}
Unlike all the stars above, HD 126516 \citep{Fekel_2019} does not contain a pulsating primary star. It contains an F-type star in a circular eclipsing binary with an M-dwarf. A third body that is likely a K or M-dwarf is detected via radial velocity measurements of the inner F star. The primary's spin rate is $v \sin i \simeq 4 \pm 1 \, {\rm km/s}$, which is about ten times lower than the expected value for an aligned and synchronous rotation. We compute the corresponding spin period in Table \ref{tab:system_params} using the measured $v \sin i$, radius, and an inclination distribution ranging from $30^{\rm o} < i < 90^{\rm o}$.

\subsection{Derived Constraints}

Using the measurements for each system, we now assume that the stars' sub-synchronicity is due to the systems being captured in CS2, and predict the properties of the outer star. We first calculate the expected value of $\eta_{\rm sync}$ if the system is in CS2 (assuming $\eta \ll 1$), for equilibrium tidal theory:
\begin{equation}
\label{eq:eta_sync}
    \eta_{\rm sync} \simeq
    \bigg( \frac{\textnormal{P}_{\rm orb}}{\textnormal{P}_{\rm rot}} \bigg)^2
    \frac{1}{2 \cos{I}} \,
    \textnormal{,}
\end{equation}
We then calculate the semi-major axis of the outer orbit via a re-arrangement of equation \ref{eq:eta}:
\begin{equation}
\label{eq:aout}
    a_{\rm out} = \Bigg(\frac{3 k}{2 k_2}\frac{M_{\rm out}M_1}{M_2(M_1+M_2)} \frac{a^6}{R^3} \frac{\cos{I}}{\eta_{sync}} \Bigg)^{1/3} 
    \textnormal{.}
\end{equation}
The projected radial velocity of the inner binary about the tertiary system's center of mass is
\begin{equation}
    K_{12} = m_{\rm out} \sin i \sqrt{\frac{G}{a_{\rm out} (m_{\rm out} + M_1 + M_2)}} \, .
\end{equation}


For the mutual inclination, we consider the most probable values, $I = 60.0^{+27.1}_{-41.8}$, using the mean and a 90\% confidence interval for a uniform distribution. For $k_2$ and $k$, we interpolate in mass and radius between values given in \cite{Claret_2019}. For the ratio $3 k/k_2$ we find: KIC 4142768: 39.6, KIC 9850387: 31.4, KIC 4480321: 38.8,
KIC 8429450: 40.3, KIC 8197761: 36.1, 
HD 201433: 31.0.

\begin{figure}
\centering
\includegraphics[scale=\scale]{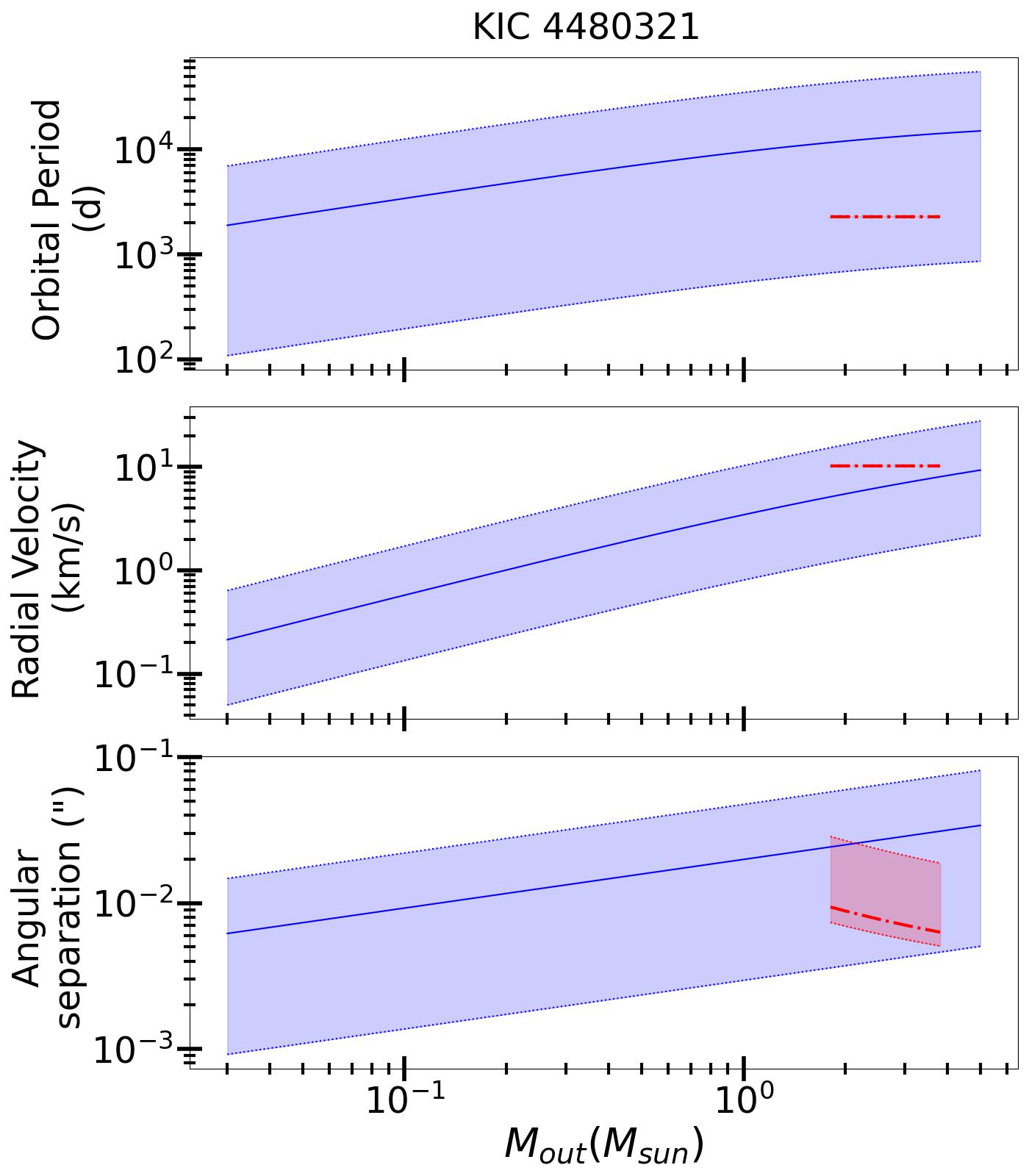}
\caption{Constraints on an outer body in the KIC 4480321 system as a function of its mass $M_{\rm out}$.
{\bf Top:} The range in predicted orbital period of the tertiary star. The blue line shows predicted values based on the observed spin period, assuming the system is in CS2 (equation \ref{eq:aout}). The red line shows the observed value for the range of possible tertiary companion masses. Blue and red regions show confidence intervals including uncertainties in the system properties. {\bf Middle:} Same as top panel, but for the radial velocity of the binary about the center of mass with the outer companion. {\bf Bottom:} The angular separation between the inner binary and outer tertiary, given the distance to the system. The predicted value is $a_{\rm out}/D$, while the observed value is measured along the line of sight, $a_{\rm out} \sin i/D$.}
\label{fig:44_constraints}
\end{figure}

\begin{figure}
\centering
\includegraphics[scale=\scale]{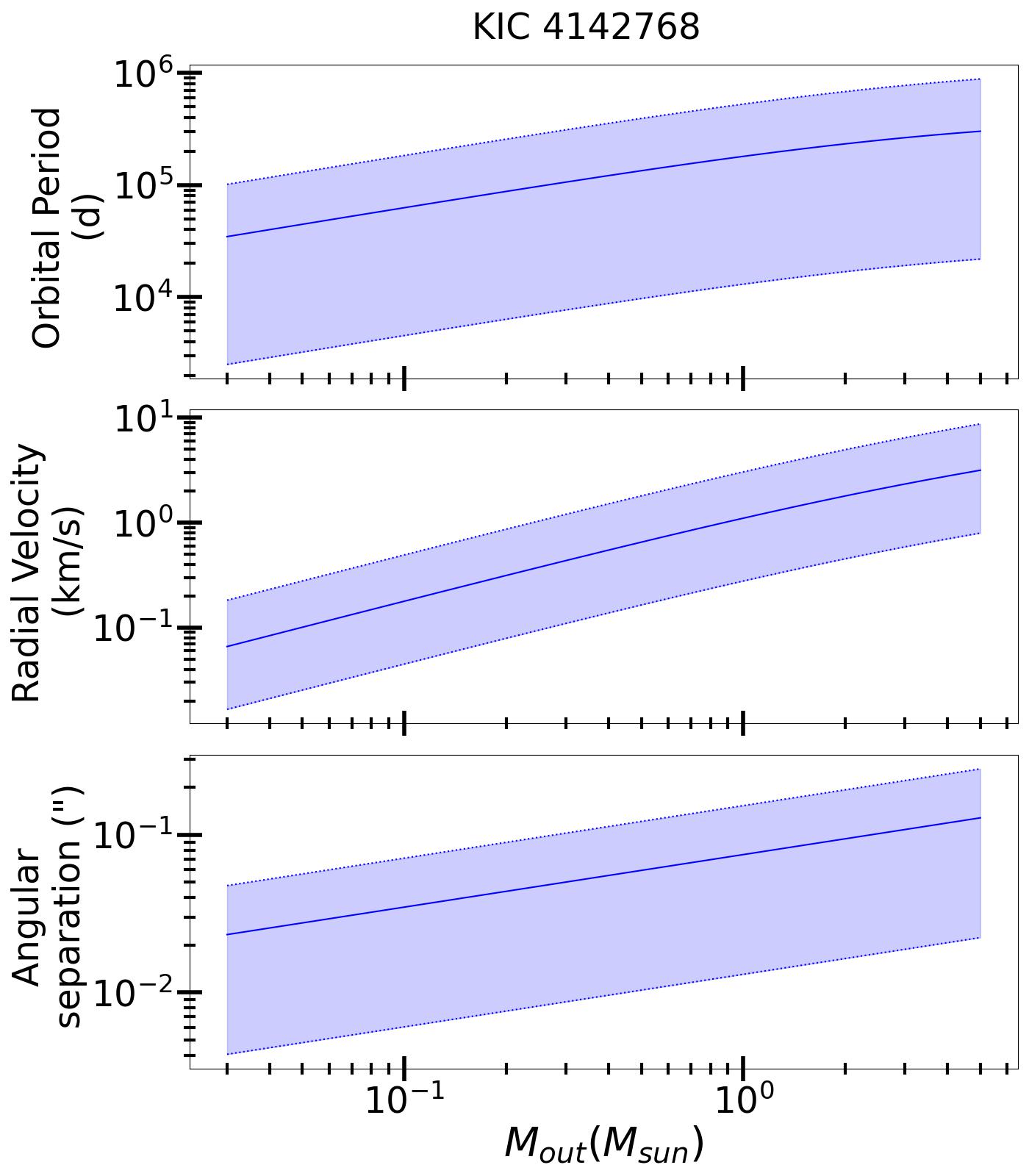}
\caption{Same as Figure \ref{fig:44_constraints} for an outer body in the KIC 4142768 system. In this case, there are no published observational constraints on the tertiary.}
\label{fig:41_constraints}
\end{figure}

\begin{figure}
\centering
\includegraphics[scale=\scale]{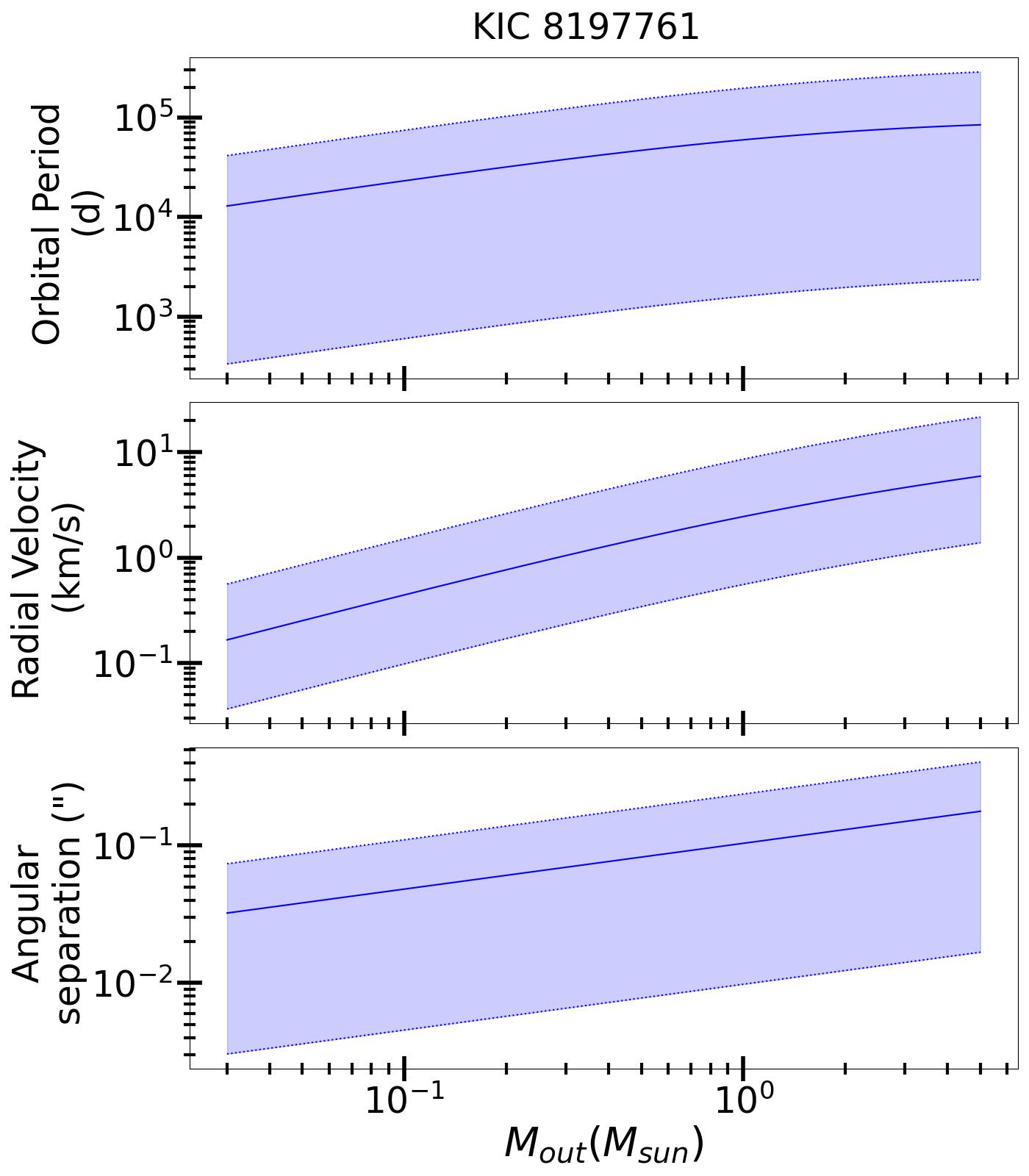}
\caption{Same as Figure \ref{fig:44_constraints} for an outer body in the KIC 8197761 system.}
\label{fig:81_constraints}
\end{figure}

\begin{figure}
\centering
\includegraphics[scale=\scale]{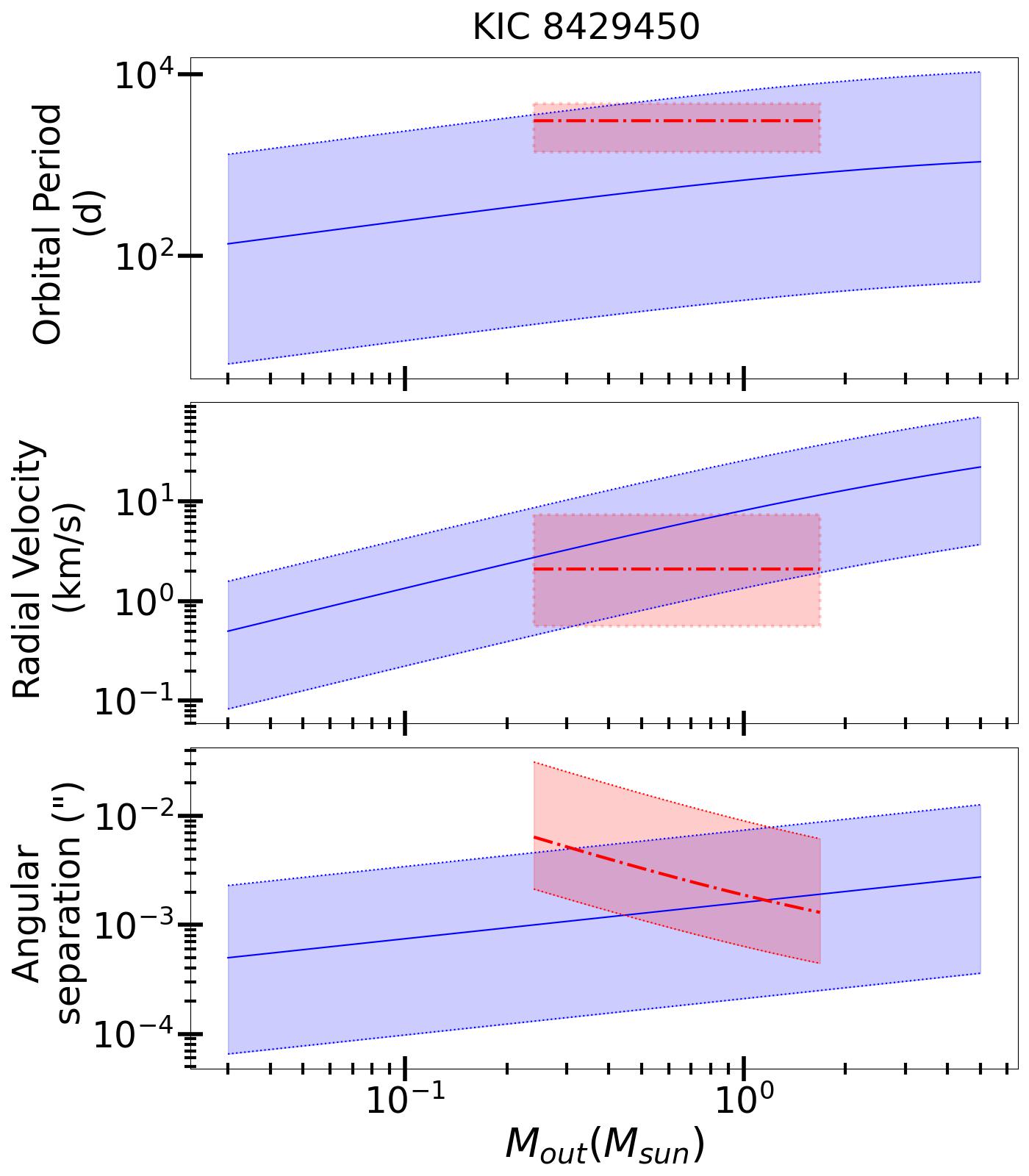}
\caption{Same as Figure \ref{fig:44_constraints} for an outer body in the KIC 8429450 system.}
\label{fig:84_constraints}.
\end{figure}

\begin{figure}
\centering
\includegraphics[scale=\scale]{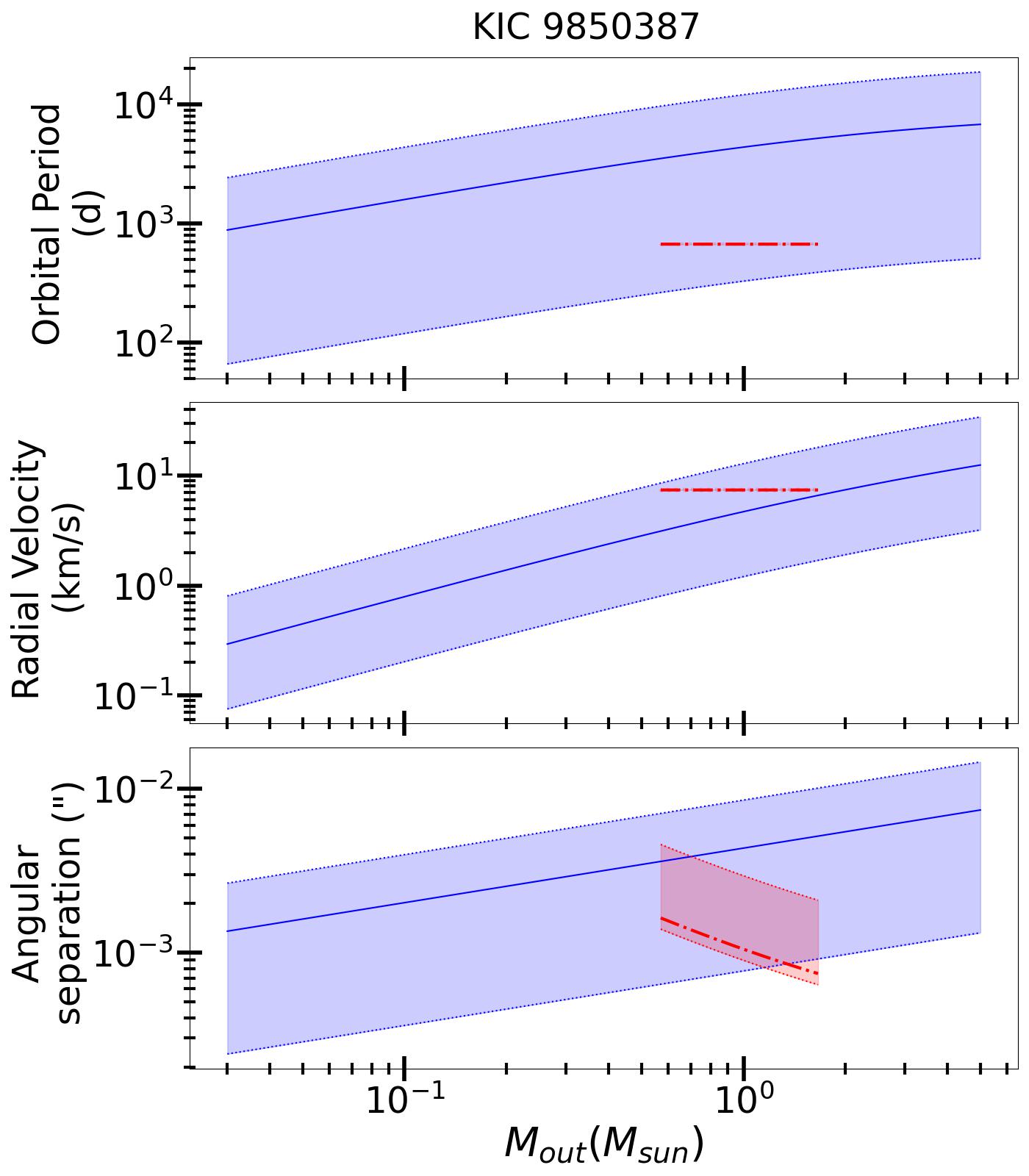}
\caption{Same as Figure \ref{fig:44_constraints} for an outer body in the KIC 9850387 system.}
\label{fig:98_constraints}
\end{figure}

\begin{figure}
\centering
\includegraphics[scale=\scale]{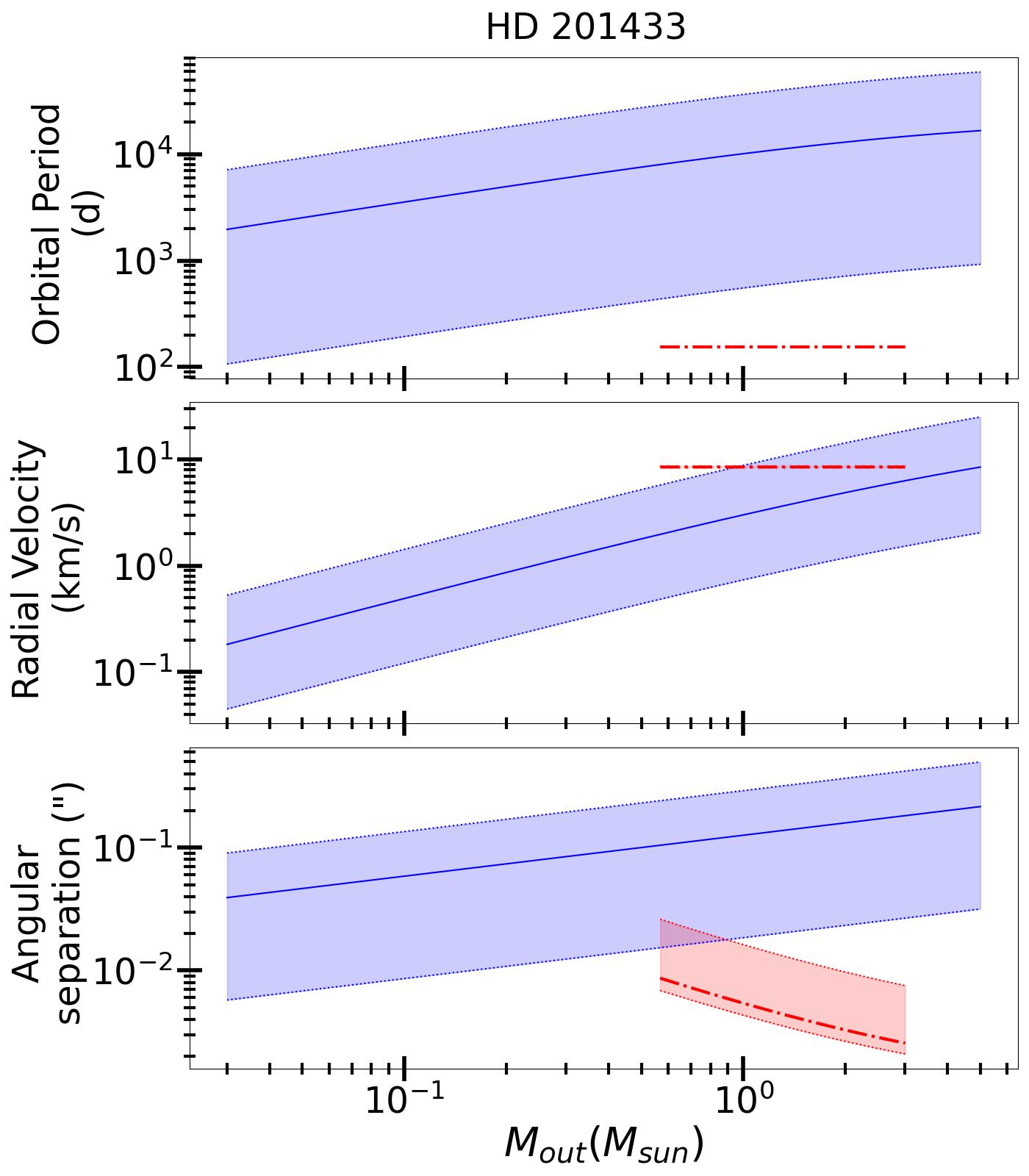}
\caption{Same as Figure \ref{fig:44_constraints} for an outer body in the HD 201433 system.}
\label{fig:20_constraints}
\end{figure}

\begin{figure}
\centering
\includegraphics[scale=\scale]{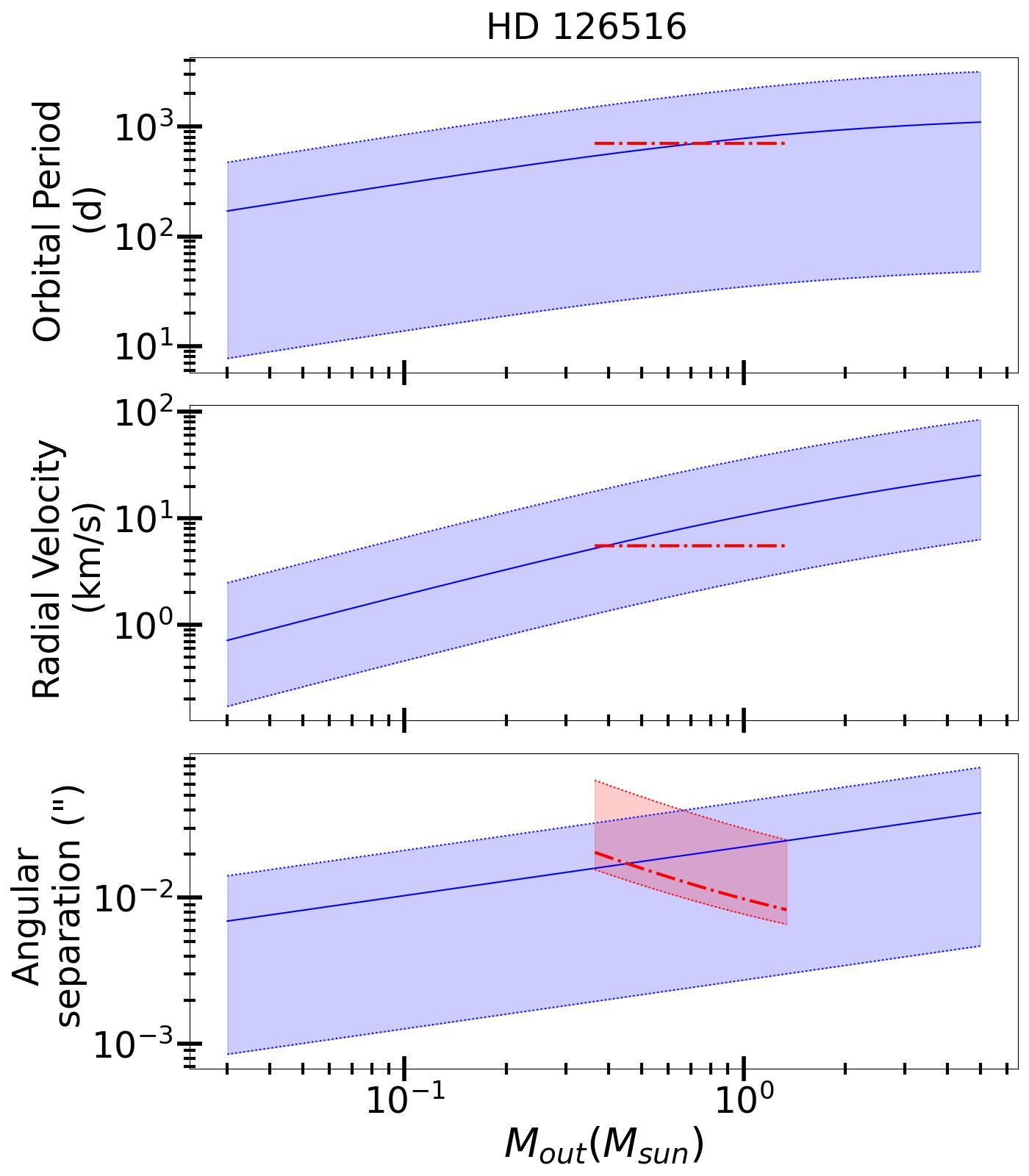}
\caption{Same as Figure \ref{fig:44_constraints} for an outer body in the HD 126516 system.}
\label{fig:12_constraints}
\end{figure}

The results of these calculations are shown in Figures \ref{fig:44_constraints}-\ref{fig:12_constraints}. We plot the expected outer orbital period ($P_{\rm out}$), radial velocity of the inner binary about the system's center of mass ($K_{12}$), and angular separation of the outer tertiary, as a function of the mass of the tertiary star. For each case, shaded blue regions span the range of possible values using the uncertainty range of the measured quantities.

For the known triple systems, we can check consistency of our results with observational constraints. In Figures \ref{fig:44_constraints}-\ref{fig:12_constraints}, the red bars/lines indicate the measured properties of the outer orbit, when available. We use the values shown in Table \ref{tab:system_params} to calculate these values. The outer tertiaries have all been measured via RVs or eclipse timing variations, and the latter translates directly RVs given the outer orbital period via
\begin{equation}
K_{12} = \frac{2 \pi a_{12} \sin i}{P_{\rm out}}
\end{equation}
where $a_{12} \sin i$ is measured via eclipse timing.

The angular separation between the inner binary and outer tertiary can potentially be measured via direct imaging of the third body. The outer semi-major axis is related to the measured properties as
\begin{equation}
\label{eq:aoutrv}
a_{\rm out} = \frac{M_1 + M_2 + M_{\rm out}}{M_{\rm out}} \frac{K_{12} P_{\rm out}}{2 \pi \sin i} \, .
\end{equation}
The corresponding angular separation is then computed given the distance to each system. In Figures \ref{fig:44_constraints}-\ref{fig:12_constraints}, the red shaded area indicates the possible range in angular separation computed via equation \ref{eq:aoutrv}.

In addition, there is a minimum outer companion mass that can be computed from the mass function,
\begin{equation}
    M_{\rm out}^3 \sin^3 i = \frac{(M_1 + M_2 + M_{\rm out})^2 P_{\rm out} K_{12}^3}{2 \pi G} \, .
\end{equation}
We assume a maximum tertiary mass of $M_1$ for all systems apart from KIC 4480321, where the bright tertiary dominates the spectrum and appears to be less than $3.5 \, M_\odot$ \citep{Lampens_2018}. 

For two of the systems (KIC 4480321, KIC 9850387), the observed outer orbital period is shorter by a factor of a $\sim 10$ compared to the most probable values. Similarly, the RVs of the outer orbits are larger than expected by a factor of a few, and the angular separations are smaller by a factor of several. This problem also exists and is most severe for HD 201433, whose observed orbital period is smaller by a factor of nearly 100 compared to what is expected.

This issue was also pointed out in Fuller \& Felce in prep, though they did not perform detailed analyses of each system. The problem can be equivalently stated that the observed stellar rotation periods are slower than expected given the known locations of tertiaries in those systems. A long stellar rotation period entails a small value of $\eta_{\rm sync}$ (equation \ref{eq:eta_sync}), and hence a large outer orbital semi-major axis (equation \ref{eq:aout}) and orbital period.


For HD 126516, the observed tertiary orbital period and inner binary RV are consistent with the predicted values. For KIC 8429450, the observed tertiary orbital period is roughly consistent (albeit slightly longer) with what is expected. For the systems without observational constraints on tertiaries, we can predict the range of expected outer orbital periods by assuming a companion mass of $0.1\textsf{-}1 \textnormal{M}_{\rm \odot}$. This yields estimates of $P_{\rm out} \sim 5 \times 10^3 \textsf{-} 5 \times 10^5$ days for KIC 4142768 and $P_{\rm out} \sim 10^3 \textsf{-} 10^5$ days for KIC 8197761. However, we caution that the eccentric orbit of KIC 4142768 will change the spin and orbital precession rates, which should be factored into more accurate predictions.



\def \gapsize {-0.25in}

\subsection{Orbital Trajectories and Capture Probability}

Since KIC 4480321 is the triple with one of the best agreements between predicted and measured tertiary properties, we calculate examples of its prior orbital evolution. We use the properties in Table \ref{tab:system_params} and calculate $\eta_{\rm sync}$ using equation \ref{eq:eta_sync}. We use equilibrium tidal theory with $g t_s = 100$. We then integrate the equations of motion: Equations \ref{eq:th}, \ref{eq:phi}, and \ref{eq:omega}.

\begin{figure}
\centering
\hspace{\gapsize}
\includegraphics[scale=\scale]{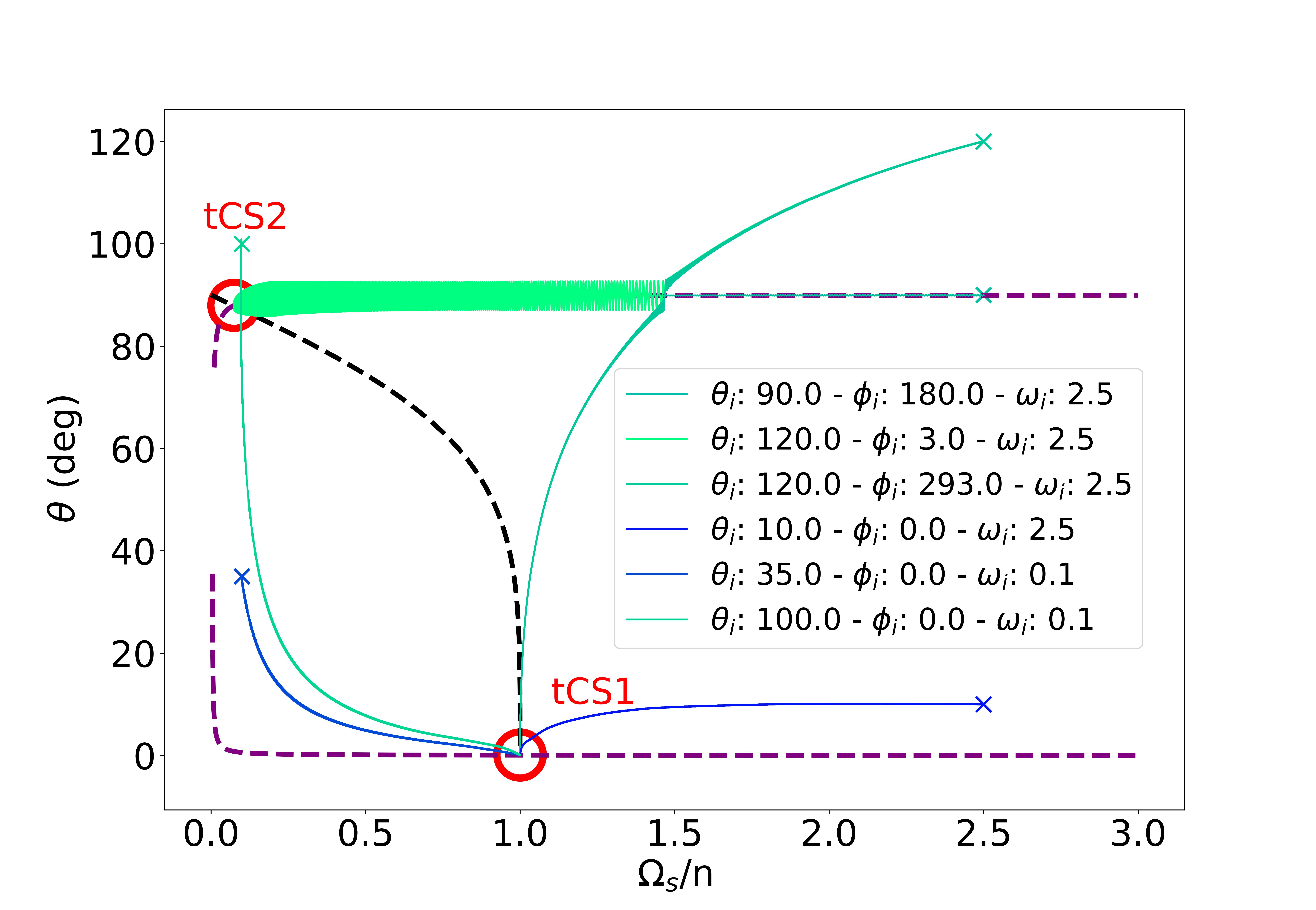}
\caption{Phase-space trajectories for KIC 4480321, similar to Figure \ref{fig:mag_traj}, showing convergence to either Cassini state 1 or 2. Initial conditions for each trajectory are indicated in the legend. For this system $\eta_{\rm sync} = 0.0029$, and we used $|g|t_s=100$. Since $\eta_{\rm sync}$ is quite small, most initial conditions converge to CS1, with only a small fraction captured into CS2.}
\label{fig:44_traj}
\end{figure}

The results are shown in Figure \ref{fig:44_traj}, for several different choices of initial conditions. For small values of $\eta_{\rm sync}$, capture into CS2 requires an initial obliquity of around or above $90^{\circ}$. Using the analytic expression given in \cite{Su_2021}, the probability capture into CS2 is
\begin{equation}
    P_{\rm CS2} \approx \frac{4 \sqrt{\eta_{\rm sync} \sin{I}}}{\pi}
    \Bigg[ \sqrt{\frac{n}{\Omega_{s,i}}} + \frac{3}{2(1+\sqrt{n/\Omega_{s,i}})} \Bigg]
    \textnormal{,}
\end{equation}
for an initial spin rate $\Omega_{s,i}$. Taking an initial spin rate of $\Omega_{s,i} = 3n$, corresponding to an initial spin period of $P_{\rm spin} \sim 3 \, {\rm d}$, we arrive at probabilites between $\sim3-10\%$, depending on the mutual inclination of the orbits.

The other known triples in this sample also have small values of $\eta_{\rm sync}$ ranging from $\sim 10^{-3}-1$. hence the capture probability into CS2 is typically quite small. This may explain the rarity of such systems relative to those in CS1 which are nearly synchronized and aligned.
 Evolution into CS2 can occur through tidal dissipation alone, as in Figure \ref{fig:44_traj}. However, other mechanisms such as changing stellar properties during early evolution (e.g., \citealt{Anderson_2018}), might also contribute to capture into tCS2.

\section{Investigation of Sub-synchronous Solar-type Binaries}
\label{sec:lurie}

While the systems discussed above are intermediate-mass stars, we also expect lower mass stars to be able to enter CS2. \cite{Lurie_2017} investigate 816 EBs composed primarily of FGK-type main sequence star primaries, using starspot modulations to deduce the rotation period of the primary star in each system. Among short-period systems ($P \lesssim 10 \, {\rm days}$), most stars are nearly tidally synchronized. They also find a population of binaries with slightly sub-synchronous rotation rates (on average 13\% slower than their orbital period). In addition, their sample contains a population of potentially very sub-synchronous rotators (25 out of 816 systems) in short-period orbits ($P \! <$10 days) with a rotation rate less than half of the orbital frequency. We investigate these systems to determine whether this sub-synchronicity can be explained via the CS2 equilibrium.

\begin{figure}
\centering
\hspace*{-0.2in}
\includegraphics[scale=0.25]{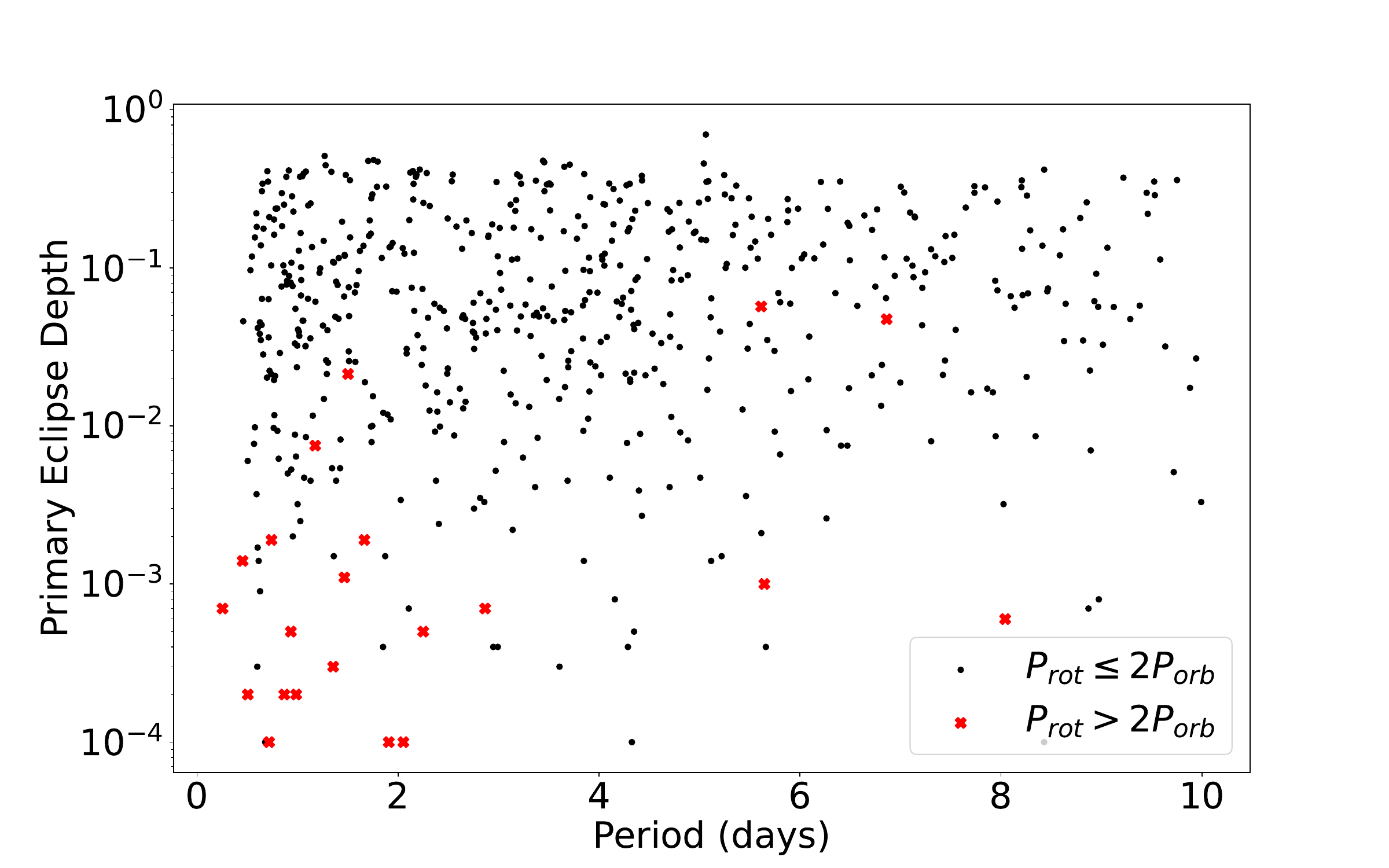}
\caption{Primary eclipse depths against orbital periods for the short-period ($P<10 \, {\rm d}$) systems cataloged in \citealt{Lurie_2017}. The very sub-synchronous systems are shown in red and clearly have shallow eclipse depths, suggesting that most contain planets or blended binaries.}
\label{fig:eclipse_depths}
\end{figure}

When examining the 25 very sub-synchronous systems, we noticed that nearly all of them lie near the bottom end of the distribution of primary eclipse depths, i.e., they have shallow transits. This is unlikely to occur by chance, and is shown in Figure \ref{fig:eclipse_depths}.
Since shallow transits could indicate an eclipsing planet rather than a binary star system, we cross-referenced each system with the NASA exoplanet catalog\cite{KeplerFP}.
Of the 25 systems, 24 are listed in the NASA exoplanet false positive list. Of these, 4 are listed as not examined, and 4 are listed as possible planets. These systems could be genuine planetary transits, in which case the tidal spin-up of the host star occurs on long time scales due to the planet's low relative mass. Rather than being in CS2, the star's slow rotation is most likely due to a lack of tidal spin-up.

Out of the 16 certified false positives, 11 are categorized as likely to be eclipsing binaries by the Kepler False Positive Working Group (\citealt{KeplerFP}). In these cases, the shallow observed eclipse depths can be explained if these systems are blended eclipsing binaries, where a background eclipsing binary is on the same pixel as a bright foreground star, or a bright tertiary companion. The measured spin rate would most likely be that of the foreground star, rather than either star in the binary.

Using the false positive probabilities listed (see \citealt{Montet_2015}), we compare the probabilities of different binary scenarios, listed in Table \ref{tab:probs}. Un-blended binaries (UEB) describe the systems we are looking for, where the observed eclipse is that of the primary star, with shallow transits produced by grazing eclipses. Blended binaries (BEB) are systems where the primary star (which produces the observed spot modulation) is not in a binary system, but is bleneded with light from a foreground/background eclipsing binary. Hierarchical-triple eclipsing binaries are denoted by HEB. Another possible scenario is a double period system (dbl), where similar primary and secondary eclipses are mistaken for two primary eclipses, resulting in the mis-classification of the system as a low mass ratio system with half the actual period. FP denotes total probability that the system represents a false positive, i.e., some type of stellar binary rather than a planetary transit.  We look for systems where the total probability to be either an un-blended binary or a hierarchical triple
is higher than the total probability to be a blended binary. This search yields only six systems, which are shown in bold in Table \ref{tab:probs}.

KIC 8938628 has a high chance of being either an un-blended binary or a hierarchical triple system, while KIC 3848972 and KIC 3443790 have high chances of being un-blended binaries. These three systems are good candidates for being in CS2. Two systems are most likely double period eclipsing binaries (KIC 4456622 and KIC 10613718). KIC 4456622 would not be a candidate example of a tidal Cassini equilibrium, since its corrected orbital period would not satisfy our criterion for very sub-synchronous rotation. However, KIC 10613718 is highly sub-synchronous and would still qualify. KIC 11560037 could be in CS2, but it has a substantial chance of being a planetary transit (low FP). 

\begin{table*}
\caption{Probable nature of sub-synchronous binaries likely to be eclipsing binaries. The systems which are more likely to be un-blended binaries or hierarchical triples than blended binaries are shown in bold, and these systems are the best candidates for stars caught in CS2. Labels are HEB: hierarchical triple, UEB: un-blended binary, BEB: blended binary, 'dbl': orbital period twice that recorded, FP: false positive.}
\centering
\begin{tabular}{@{}cccccccccc@{}}
\toprule
\textbf{KID} & 
\textbf{HEB} &\textbf{UEB} & \textbf{BEB} & \textbf{HEB} &
\textbf{UEB} &
\textbf{BEB} &
\textbf{FP} & $\textbf{P}_{\boldsymbol{\rm orb}}$ &  $\textbf{P}_{\boldsymbol{\rm spin}}$ \\
 & & & & \textbf{dbl} & \textbf{dbl} & \textbf{dbl} & & & \\
 \midrule \\
$ \boldsymbol{3443790}$ & 0.051 & $ \boldsymbol{0.67}$ & 0.096 & 0.01 & 0.18 & 0.00049 & 1.0 & 1.665784 & 5.211\\
$ \boldsymbol{3848972}$ & 0.02 & $ \boldsymbol{0.77}$ & 0.14 & 0.0022 & 0.061 & 0.0066 & 1.0 & 0.741057 & 21.578\\
$ \boldsymbol{4456622}$ & 1.7e-11 & 1.1e-05 & 2.3e-07 & 0.016 & $ \boldsymbol{0.95}$ & 0.024 & 0.99 & 1.502816 & 5.721\\
4929299 & 7.6e-10 & 2.5e-06 & 0.0024 & 2.3e-05 & 0.00019 & 2.9e-05 & 0.0027 & 0.868719 & 23.764\\
4946584 & 0.0014 & 0.0058 & 0.81 & 0.0019 & 0.051 & 0.14 & 1.0 & 0.453972 & 2.329\\
5642620 & 9.2e-06 & 0.00022 & 0.12 & 3.1e-05 & 7e-05 & 0.018 & 0.14 & 0.986791 & 23.622\\
$ \boldsymbol{8938628}$ & $ \boldsymbol{0.42}$ & 0.42 & 0.0013 & 0.011 & 0.086 & 1.5e-07 & 0.94 & 6.862216 & 20.413\\
9266285 & 0.0034 & 2.6e-59 & 0.38 & 0.016 & 0.14 & 4.7e-05 & 0.55 & 5.61387 & 33.613\\
$ \boldsymbol{10613718}$ & 5.2e-07 & 0.017 & 0.00028 & 0.0014 & $ \boldsymbol{0.98}$ & 0.00087 & 1.0 & 1.175878 & 13.078\\
$ \boldsymbol{11560037}$ & 0.013 & $ \boldsymbol{0.28}$ & 0.0019 & 0.0047 & 0.033 & 0.0039 & 0.33 & 1.466522 & 20.413\\
12255382 & 1.7e-52 & 0.0 & 1.1e-14 & 1.8e-17 & 7.3e-18 & 0.88 & 1.0 & 0.999323 & 12.812 \\
\bottomrule
\end{tabular}
\label{tab:probs}
\end{table*}




\section{Discussion}
\label{sec:disc}

\subsection{Slow observed rotation rates}

We have identified several short-period stellar binaries (Table \ref{tab:system_params}) with a very slowly rotating primary star, which are good candidates to be trapped in the sub-synchronous CS2 state.
Previously, \cite{Fuller_2021} postulated that an ``inverse tidal" process could produce either very slowly rotating or very rapidly rotating stars in close binaries containing pulsating primaries. We believe this hypothesis is less likely than the CS2 hypothesis because a) one of the observed systems and the new candidates from Section \ref{sec:lurie} do not contain pulsating stars b) five out of seven of the systems in Table \ref{tab:probs} are already known to have tertiaries at about the right separation, c) we are not aware of a population of very rapidly rotating stars in close binaries as predicted by the inverse tides theory, and d) the inverse tidal process can only act over a narrow range of binary orbital period, whereas the observed systems span a range of 2-14 days.

However, there are problems with the CS2 hypothesis. Three of the observed systems have primaries that rotate slower than predicted by the weak tides assumption, suggesting additional physical mechanisms may be at work. One possibility we have explored is magnetic braking, which slows the CS2 equilibrium rotation period by a factor of $(1+t_{\rm s}/t_{\rm mag})^{-1/2}$ (equation \ref{eq:omseqmb}). However, this is unlikely to be the explanation because all but one of the systems in Table \ref{tab:system_params} are hot stars ($T_{\rm eff} \gtrsim 6500 \, {\rm K}$) that are $\gamma$-Doradus or SPB pulsators, which typically rotate rapidly and are not magnetically braked.

Even if magnetic braking does act, from equation \ref{eq:max_braking} we have a minimum magnetic braking time scale of $t_{\rm mag} \! \sim \! 150$ Myr for Sun-like stars at spin periods of 5 days. Evaluating equation \ref{eq:ts}, the tidal spin-up time scale is:
\begin{equation}
    t_s \simeq 200 \, {\rm Myr} \frac{Q}{10^6} \bigg( \frac{P_{\rm orb}}{5 \, {\rm d}} \bigg)^{3}  \, 
\end{equation}
for Sun-like stars with equal mass companions. Hence, we expect $t_{\rm mag} < t_s$ only for relatively long-period systems ($P \gtrsim 5$ d). Using a non-saturated braking prescription with $t_{\rm mag} \propto P_{\rm s}^2$ would further reduce the importance of magnetic braking at long spin periods, so this effect is likely unimportant for the stars in our sample.

We showed that dissipation via gravity waves slightly increases the expected rotation rate in CS2, so this cannot explain the very slow rotation rates. In contrast, dissipation via inertial waves can substantially reduce the expected rotation rate (see also Fuller \& Felce, in prep). The problem with this explanation is that the stars in our sample are hot stars with extremely thin surface convective zones, decreasing the importance of inertial wave dissipation in the envelope. These stars do have convective cores, which may help contribute to inertial wave dissipation, which should be studied in future work.

\subsection{Lack of low-mass stars in Cassini States}

As shown in Section \ref{sec:lurie}, there are only $\sim$4-6 good candidates for stars in CS2 in the sample of \cite{Lurie_2017}. It is remarkable to find only a few candidates among the several hundred short-period binaries of that sample. For comparison, roughly 4 out of the 35 $\gamma-$Doradus systems in close binaries in \cite{Li_2020a} have very sub-synchronous rotation indicative of being in CS2. Here we explore possible explanations.

One possibility is that Sun-like stars experience more efficient tidal dissipation that accelerates orbital decay. As discussed in Fuller \& Felce in prep, systems typically only remain in CS2 for a fraction of their main sequence lifetime, because ongoing tidal dissipation shrinks the orbit until CS2 destabilizes and the system transitions to the standard CS1 equilibrium with nearly synchronous rotation. Indeed, many recent works have investigated the eccentricity or obliquity of stellar binaries and exoplanets \citep{Winn_2010,VanEylen_2016,Bashi_2023}, finding that tidal dissipation is more efficient in cool stars ($T \lesssim 6200 \, {\rm K}$) relative to hot stars. In this case, cool stars would be caught in CS2 for a smaller fraction of their lifetimes, making it less likely to observe them in this state.

Another possibility is that there are selection effects in these samples that prevent us from finding cooler stars in CS2. Since rotation periods in \cite{Lurie_2017} are measured via spot modulation, and spot modulation becomes weaker for slowly rotating stars \citep{Cao_2022}, it may be difficult to measure the rotation periods of very slowly rotating stars in CS2. It would be worthwhile to further investigate those systems that do not show clear spot modulation, e.g., with spectroscopic measurements of $v \sin i$ to constrain the stellar rotation rate. Indeed, this is how HD 126516 \citep{Fekel_2019} was discovered, and so future spectroscopic measurements of short-period binaries could yield many more CS2 candidates.

It could also be possible that cooler stars are less likely to have tertiary companions capable of locking the inner system in CS2. We disfavor this possibility because it is known that the tertiary fraction of short-period binaries is high \citep{Tokovinin_2006}, even for cool stars. Moreover, it seems unlikely this effect could create a large difference between the $\sim \! 1 \, M_\odot$ stars of \cite{Lurie_2017} and the $\sim \! 1.5 \, M_\odot$ stars of \cite{Li_2020a}, since multiplicity fractions only vary slightly over that mass range \citep{Moe_2017}.

\subsection{Secondary Star Properties}

While our analysis has centered around the rotation rate of primary stars, it is worth commenting on some properties of the secondary star.

Firstly, although the spin rotation rate of the secondary star has usually not been measured, it is possible for the secondary star to have also been captured into CS2. Since, to first approximation, the stars within the binary have uncoupled spin-orbit dynamics, their spin rates should be independent. This would allow either or both stars to be caught in CS2.

Secondly, we consider the effect of a very low-mass secondary on the system dynamics. Since the tidal synchronization time is inversely proportional to the mass of the secondary (equation \ref{eq:ts}), systems with low-mass secondary stars are less likely to reach CS2 within their lifetime. In addition, all of our analysis relies on the assumption that the inner binary orbital angular momentum is larger than that of either star's spin. This requires:

\begin{equation}
    k M_1 R_1^2  \Omega_s \ll \frac{M_1 M_2}{M_1 + M_2} \sqrt{ G(M_1 + M_2) a} \\
    \textnormal{.}
\end{equation}
Taking the limit $M_1 \gg M_2$, and using $\Omega_s = \Omega_{\rm orb}$, we find:
\begin{align}
    M_2 > M_1 k \bigg(\frac{R_1}{a}\bigg)^2 \,.
\end{align}
For typical values $P_{\rm orb} = 3 \, {\rm d}$, $R_1 = 2 R_\odot$, $M_1 = 1 M_\odot$, $k = 0.1$, we find $M_2 \geq 0.005 M_\odot$. This suggests that the mechanism discussed above would not be applicable to systems with planetary-mass secondaries, but it could still be applicable to those with brown dwarf secondaries.

\section{Conclusion}

In this work, we investigated the possibility that slowly rotating stars in close binaries result from capture into Cassini State 2 (CS2), an equilibrium state caused by orbital precession from a tertiary star. In CS2, the primary star can have a very slow rotation and highly misaligned rotation axis relative to the inner binary orbit.  We examined the behavior of Cassini states incorporating the effects of magnetic braking and dynamical tides. We build on the results from \cite{Su_2021} and find that magnetic braking reduces the spin rate for both tidal Cassini states 1 and 2, and moves the obliquities of these two equilibria closer to each other. At a critical value of the magnetic braking strength, tidal Cassini State 1 is de-stabilized, but this is unlikely to be realized based on estimates of tidal spin-up and magnetic braking time scales.

We also considered the effect of gravity and inertial waves on the CS2 equilibria, finding slightly increased rotation rates for gravity wave dissipation, and potentially greatly decreased rotation rates when accounting for inertial wave dissipation. Large amounts of inertial wave dissipation can shift the CS2 spin-orbit misalignment away from $\theta \simeq \pi/2$ and towards $\theta = I$ or $\theta=\pi-I$, where $I$ is the inclination of the outer tertiary relative to the inner orbit.

Comparing to observed systems shown in Table \ref{tab:system_params}, we find that many have longer outer orbital periods than expected from equilibrium tidal theory (or alternatively, they have longer stellar rotation periods than expected given the observed outer orbital period). Since magnetic braking is likely unimportant for those systems which contain hot primaries ($T_{\rm eff} \gtrsim 6500 \, {\rm K}$), dissipation via inertial waves is our favored explanation. However, these stars have convective cores and only very thin convective envelopes, so it is not clear that inertial wave dissipation can be effective. We recommend further investigation into the effects of inertial waves in these kinds of systems.

We also search the eclipsing binary catalog of \cite{Lurie_2017} for systems with highly sub-synchronous rotation, finding 25 candidates. However, closer inspection of these objects revealed that most have very shallow eclipse depths, entailing they likely contain blended systems or planetary companions. We found only 4 good candidates to be in CS2, a remarkably small number relative to the sub-synchronous rotators in the asteroseismic catalog of \cite{Li_2020a}. It is possible that more efficient tidal dissipation in Sun-like stars causes their orbits to quickly decay to short periods where CS2 is destabilized, making us less likely to observe them in CS2.

The CS2 hypothesis can be tested by searching for tertiary companions, which are expected to have orbital periods of $\sim$1-100 years. Figures \ref{fig:44_constraints}-\ref{fig:12_constraints} show the predicted tertiary orbital properties as a function of tertiary mass for systems suspected to be in CS2. These types of tertiaries can be detected through either long-term RV measurements, or high-resolution direct imaging. We also recommend future spectroscopic $v \sin i$ measurements of stars in short-period binaries to search for slow rotators that could be in CS2.

\section*{Acknowledgments}

We thank Yubo Su for useful discussions and feedback on this work. 

\section{Data Availability}

The orbital integration code, and plotting scripts to make
Figures 1-13, are available upon request.

\bibliography{main}

\end{document}